\documentclass[11pt]{article}
\newcommand{\Comment}[1]{{}}
\usepackage{amsfonts,amsmath,epsfig,color,latexsym,graphics}
\usepackage{cite}
\usepackage[textwidth = 430 pt, textheight = 630 pt]{geometry}

\Comment{\usepackage{color}
\definecolor{MyDarkBlue}{rgb}{0.15,0.15,0.45}
\usepackage[linktocpage=true]{hyperref}
\hypersetup{
colorlinks=true,
citecolor=MyDarkBlue,
linkcolor=MyDarkBlue,
urlcolor=MyDarkBlue,
pdfauthor={Stephen G. Naculich, Horatiu Nastase and Howard J. Schnitzer},
pdftitle={Applications of Subleading-color
Amplitudes in N=4 SYM Theory },
pdfsubject={hep-th}
}

\usepackage[numbers,sort&compress]{natbib}
\usepackage{hypernat}}

\newcommand\ignore[1]{}
\def\one{{\,\hbox{1\kern-.8mm l}}}

\def\ket#1{\left| #1\right\rangle}

\def\Tr{{\rm Tr\, }}

\def\id{\textrm{id}}

\def\a{\alpha}\def\b{\beta}

\def\e{\epsilon}

\def\ib{{\ov i}}

\def\Z{\mathbb{Z}}

\newcommand{\Cset}{{\,\,{{{^{_{\pmb{\mid}}}}\kern-.45em{\mathrm C}}}}}

\newcommand{\cA}{\mathcal A}
\newcommand{\cC}{\mathcal C}

\newcommand{\cG}{\mathcal G}

\newcommand{\cN}{\mathcal N}\newcommand{\cO}{\mathcal O}

\def\rmcp{ {\rm c.p.}}

\def\cyclic{ {\rm cyclic} }

\def\lSYM{\lambda_{\rm SYM}}
\def\lSG{\lambda_{\rm SG}}
\def\Asy{A_{\rm SYM}}
\def\Msy{M_{\rm SYM}}
\def\Asgzero{A_{\rm SG}^{(0)}}
\def\Asgone{A_{\rm SG}^{(1)}}

\def\Msgone{M_{\rm SG}^{(1)}}
\def\Msgtwo{M_{\rm SG}^{(2)}}
\def\MsgEll{M_{\rm SG}^{(L)}}
\def\Tr{{\rm Tr}}
\def\e{ {\rm e} }

\def\half{ {1\over 2} }
\def\third{ {1\over 3} }
\def\lr{\leftrightarrow}

\def\eqn#1{eq.~(\ref{#1})} 
\def\eqns#1#2{eqs.~(\ref{#1}) and~(\ref{#2})}

\def\ket#1{|{#1}\rangle}

\def\cA{  {\cal A}  }
\def\cC{  {\cal C}  }
\def\cG{  {\cal G}  }
\def\cG{  {\cal G}  }

\def\cN{  {\cal N}  }
\def\cO{  {\cal O}  }
\def\bT{\mathbf{T}}
\def\bF{\mathbf{F}}
\def\bG{\mathbf{G}}
\def\bGam{\mathbf{\Gamma}}
\def\bone{1\kern -3pt \mathrm{l}}

\def\ii#1{ {[{#1}]} }
\def\Ell{{(L)}}
\def\Ellk{{(L,k)}}
\def\Elleven{{(L,2k)}}
\def\Ellodd{{(L,2k+1)}}
\def\EllL{{(L,L)}}
\def\Ellf{{(Lf)}}

\def\Atree{  \left( - 8 i K \over s t u \right) }
\def\Zero{{(0)}}
\def\One{{(1)}}
\def\Onef{{(1f)}}
\def\OneLC{{(1,0)}}

\def\OneDT{{(1,1)}}

\def\Two{{(2)}}
\def\Twok{{(2,k)}}
\def\Twof{{(2f)}}
\def\TwoLC{{(2,0)}}

\def\TwoDT{{(2,1)}}

\def\TwoSC{{(2,2)}}

\def\TwoP{{(2)P}}
\def\TwoNP{{(2)NP}}
\def\Three{{(3)}}

\def\Threef{{(3f)}}

\def\ep{\epsilon}

\def\tS{{\tt S}}
\def\tT{{\tt T}}
\def\tU{{\tt U}}
\def\ia{ \alpha_\ep}

\def\ib{\beta_\ep}

\def\ic{\gamma_\ep}

\def\id{\delta_\ep}

\def\ones{ \pmatrix{1\cr 1\cr 1} }

\newcommand{\be}{\begin{equation}}
\newcommand{\bea}{\begin{eqnarray}}

\newcommand{\ee}{\end{equation}}
\newcommand{\eea}{\end{eqnarray}}

\def\eps{ \epsilon }

\def\Azero{ A^\Zero} 
\def\bF{\mathbf{F}}
\def\bGam{\mathbf{\Gamma}}
\def\bG{\mathbf{G}}

\def\bone{1\kern -3pt \mathrm{l}}
\def\bS{\mathbf{S}}
\def\bT{\mathbf{T}}
\def\cA{  {\cal A}  }
\def\cC{  {\cal C}  }
\def\cG{  {\cal G}  }

\def\cN{  {\cal N}  }
\def\cO{  {\cal O}  }

\def\Ellf{{(Lf)}}
\def\Ellk{{(L,k)}}
\def\Elleven{{(L,2m)}}
\def\Ellodd{{(L,2m+1)}}
\def\Ellevenplustwo{{(L,2m+2)}}
\def\Ellzero{{(L,0)}}
\def\Ellone{{(L,1)}}

\def\Ell{{(L)}}
\def\EllL{{(L,L)}}
\def\ep{\epsilon}
\def\eps{\epsilon}
\def\eqn#1{eq.~(\ref{#1})} 
\def\eqns#1#2{eqs.~(\ref{#1}) and~(\ref{#2})}

\def\e{ {\rm e} }
\def\goell{  f_\ell}
\def\gone{ f_1}
\def\gotwo{  f_2}
\def\gzell{  g_\ell}
\def\gz{  g_1}
\def\gztwo{  g_2}
\def\half{ {1\over 2} }
\def\ia{\alpha}
\def\ib{\beta}
\def\ic{\gamma}
\def\id{\delta}

\def\ii#1{ {[{#1}]} }
\def\ket#1{|{#1}\rangle}

\def\lr{\leftrightarrow}
\def\mommu{ { s_{ij} \over \mu^2 } }
\def\mom{ { s_{ij} \over Q^2 } }
\def\MsgEll{M_{\rm SG}^{(L)}}
\def\Msgone{M_{\rm SG}^{(1)}}
\def\Msgtwo{M_{\rm SG}^{(2)}}
\def\Msy{M_{\rm SYM}}
\def\muQell{ \left( \mu^2 \over Q^2 \right)^{\ell \eps} }
\def\muQone{ \left( \mu^2 \over Q^2 \right)^{\eps} }
\def\muQtwo{ \left( \mu^2 \over Q^2 \right)^{2\eps} }
\def\One{{(1)}}
\def\Onef{{(1f)}}
\def\Onezero{{(1,0)}}
\def\Oneone{{(1,1)}}
\def\ones{ \begin{pmatrix}1\cr 1\cr 1\end{pmatrix} }

\def\pel{{(\ell)}}

\def\Qmu{ {Q^2 \over \mu^2 } }

\def\suml{\sum_{\ell=1}^\infty}
\def\sumL{\sum_{L=0}^\infty}

\def\theequation{\thesection.\arabic{equation}}
\def\third{ {1\over 3} }

\def\Tr{{\rm Tr}}
\def\tS{{\cal S}}
\def\tT{{\cal T}}
\def\tU{{\cal U}}
\def\Two{{(2)}}
\def\Twof{{(2f)}}
\def\Twok{{(2,k)}}
\def\Twozero{{(2,0)}}
\def\Twoone{{(2,1)}}
\def\Twotwo{{(2,2)}}
\def\TwoNP{{(2)NP}}
\def\TwoP{{(2)P}}
\def\Three{{(3)}}
\def\Threef{{(3f)}}

\def\Zero{{(0)}}

\begin{document}

\renewcommand{\thefootnote}{\fnsymbol{footnote}}

\makeatletter
\@addtoreset{equation}{section}
\makeatother
\renewcommand{\theequation}{\thesection.\arabic{equation}}

\rightline{}
\rightline{}
   \vspace{1.8truecm}

\begin{flushright}
BRX-TH-636\\ BOW-PH-149
\end{flushright}

\vspace{10pt}

%%%%%%%%%%%%%%%%%

\begin{center}
{\LARGE \bf{\sc Applications of Subleading-color
Amplitudes in ${\cal N}=4$ SYM Theory 
}}
\end{center} 
 \vspace{1truecm}
\thispagestyle{empty} \centerline{
{\large \bf {\sc Stephen G. Naculich${}^{a,}$}}\footnote{E-mail address: \Comment{\href{mailto:naculich@bowdoin.edu}}{\tt 
    naculich@bowdoin.edu}},
    {\large \bf {\sc Horatiu Nastase${}^{b,}$}}\footnote{E-mail address: \Comment{\href{mailto:nastase@ift.unesp.br}}{\tt 
    nastase@ift.unesp.br}} {\bf{\sc and}}
    {\large \bf {\sc Howard J. Schnitzer${}^{c,}$}}\footnote{E-mail address:
                                \Comment{ \href{mailto:schnitzr@brandeis.edu}}{\tt schnitzr@brandeis.edu}}
                                                           }

\vspace{1cm}

\centerline{{\it$^{a}$ Department of Physics}} \centerline{{\it
Bowdoin College, Brunswick, ME 04011, USA}}

\vspace{.8cm}

\centerline{{\it ${}^b$ 
Instituto de F\'{i}sica Te\'{o}rica, UNESP-Universidade Estadual Paulista}} \centerline{{\it 
R. Dr. Bento T. Ferraz 271, Bl. II, Sao Paulo 01140-070, SP, Brazil}}

\vspace{.8cm}
\centerline{{\it ${}^c$ 
Theoretical Physics Group, Martin Fisher School of Physics}} \centerline{{\it Brandeis University, Waltham, 
MA 02454, USA}}

\vspace{2truecm}

%%%%%%%%%%%%%%%%%
\thispagestyle{empty}

\centerline{\sc Abstract}

\vspace{.4truecm}

\begin{center}
\begin{minipage}[c]{380pt}{\noindent 
A number of features and applications of subleading color amplitudes of ${\cal N}=4$ SYM theory are reviewed.
Particular attention is given to the IR divergences of the subleading-color amplitudes, the relationships of ${\cal N}=4$ SYM theory to ${\cal N}=8$ 
supergravity, and to geometric interpretations of one-loop subleading color and $N^kMHV$ amplitudes of ${\cal N}=4 $ SYM theory.
}
\end{minipage}
\end{center}

\vspace{.5cm}

\setcounter{page}{0}
\setcounter{tocdepth}{2}

\newpage

%\tableofcontents
\renewcommand{\thefootnote}{\arabic{footnote}}
\setcounter{footnote}{0}

\linespread{1.1}
\parskip 4pt

{}~
{}~

\section{Introduction}

Planar amplitudes of ${\cal N}=4$ SYM theory have been extensively studied by a variety of methods, 
see e.g. refs.~\cite{Witten:2003nn,
Cachazo:2004kj,
Britto:2004nc,
Britto:2004ap,
Britto:2005fq,
Hodges:2005bf,
Bern:2005iz,
Hodges:2005aj,
Hodges:2006tw,
Alday:2007hr,
Drummond:2007aua,
Brandhuber:2007yx,
Drummond:2007cf,
Drummond:2007au,
ArkaniHamed:2008gz,
Drummond:2008cr,
Hodges:2009hk,
ArkaniHamed:2009dn,
ArkaniHamed:2009vw,
Hodges:2010kq,
Bullimore:2010pj,
Mason:2010yk,
ArkaniHamed:2010gh}.
For a recent overview see ref.~\cite{Dixon:2011xs} 
and the special issue of Journal of Physics A, devoted to ``Scattering Amplitudes in Gauge Theories."
Subleading color (i.e., non-planar) amplitudes, however,  usually receive less attention 
\cite{Catani:1996jh,Catani:1998bh,Sterman:2002qn,MertAybat:2006wq,MertAybat:2006mz,
Becher:2009cu,Gardi:2009qi,Dixon:2009gx,Becher:2009qa}. 
Nevertheless interesting insights are available from various applications 
of subleading color amplitudes. 
One case in point is a possible weak/weak duality between ${\cal N}=4 $ SYM theory and ${\cal N}=8$ supergravity 
\cite{Green:1982sw,
Berends:1988zp,
Bern:1998ug,
Bern:1998sv,
BjerrumBohr:2006gg,
ArkaniHamed:2008gz,
Naculich:2008ew,
Bern:2008qj,
Naculich:2008ys,
Bern:2010ue,
Bern:2010yg,
BjerrumBohr:2010ta,
BjerrumBohr:2010yc,
BjerrumBohr:2010hn}.
Since non-planar graphs appear on an equal footing with planar graphs
in ${\cal N}=8$ supergravity, one needs to understand the non-planar
graphs in ${\cal N}=4$ SYM if the weak/weak duality is to be explored.

This review will cover three significant topics. Section 2 discusses the
IR divergences of various subleading-color amplitudes. In section 3 the
interplay between subleading-color amplitudes of ${\cal N}=4 $ SYM theory
and amplitudes of ${\cal N}=8$ supergravity will be considered. Section 4
presents various geometric interpretations of one-loop subleading-color
amplitudes, primarily using the tools of momentum twistors and the
accompanying polytope interpretation.

In the remainder of this section we define the notation for the color
decomposition, the loop expansion, and the $1/N$ expansion.

At tree level, we can decompose the amplitudes ${\cal A}_n$ of ${\cal N}=4$ SYM into {\em color-ordered} tree amplitudes $A_n$ 
\bea
{\cal A}_n^{tree}(12...n)&=&g^{n-2}\sum_{\sigma\in S_n/Z_n}\Tr(T^{a_{\sigma(1)}}...T^{a_{\sigma(n)}})A_n^{tree}(\sigma(1)...\sigma(n))\cr
&=&g^{n-2}\sum_{P(23...n)}\Tr(T^{a_1}T^{a_{P(2)}}...T^{a_{P(n)}})A_n^{tree}(1P(2)...P(n))
\eea
where in the second line, 
$1$ is fixed and $P(23..n)$ is a permutation of $2,3,\ldots,n$,
and where $T^a$ are SU$(N)$ generators in the fundamental representation,
normalized according to $\Tr (T^a T^b) = \delta^{ab}$.
The color-ordered amplitudes $A_n$ depend on the momenta and 
polarizations of the external particles.

The color-ordered amplitudes are not independent. 
For $n$-point amplitudes, there is a basis of $(n-2)!$ amplitudes 
out of the total $n!$, 
called the Kleiss-Kuijf (KK) basis \cite{Kleiss:1988ne}
and we can find the others easily in terms of it \cite{Bern:2008qj}. 
It is based on the existence of the Kleiss-Kuijf relations \cite{Kleiss:1988ne}
\be
A_n(1,\{\a\},n,\{\b\})=(-1)^{n_\b}\sum_{\{\sigma\}_i\in OP(\{\a\},\{\b^T\})}A_n(1,\{\sigma\}_i,n)\label{KK}
\ee
where $\sigma_i$ are ordered permutations, 
i.e. ones that keep the order of $\{\a\}$ and of $\{\b^T\}$ inside $\sigma_i$. 
Thus the KK basis is $A_n(1,{\cal P}(2,...,n-1),n)$ where ${\cal P}$ are arbitrary permutations.
All the other $A_n$'s can be recovered from it by use of the KK relations and cyclicity and reflection invariance
\be
A_n(12...n)=(-1)^nA_n(n...21)
\ee

At one loop, we can write a similar expansion in color-ordered amplitudes,
\bea
{\cal A}_n^{1-loop}(12...n)&=&g^n \sum_{j=1}^{[n/2]+1}\sum_{\sigma\in S_n/S_{n;j}}Gr_{n;j}(\sigma)A_{n;j}(\sigma(1)...\sigma(n))\cr
Gr_{n;1}(1)&=&N \, \Tr(T^{a_1}...T^{a_n})\cr
Gr_{n;j}(1)&=&\Tr(T^{a_1}...T^{a_{j-1}})\Tr(T^{a_j}...T^{a_n})\label{nonplanar}
\eea
However, the subleading piece in the $1/N$ expansion can be obtained from the leading piece by
\be
A_{n;j}(12...,j-1,j,j+1,...n)=(-1)^{j-1}\sum_{\sigma\in COP\{\a\},\{\b\}}A_{n;1}(\sigma)\label{subleadingrel}
\ee
where $COP$ are cyclically ordered permutations, again keeping the order of $\{\a\}$ and $\{\b\}$ fixed up to cyclic permutations.

At arbitrary loops, the decomposition of the four-gluon amplitude takes a form with only single and double traces
\bea
\hskip-.4in{\cal A}_{4} (1234)
&=&g^2
\sum_{\sigma\in S_4/\Z_4}
\Tr(T^{a_{\sigma(1)}}T^{a_{\sigma(2)}}T^{a_{\sigma(3)}} T^{a_{\sigma(4)}})
N \;A_{4;1}(\sigma(1)\sigma(2)\sigma(3)\sigma(4))
\nonumber\\
&+& g^2\sum_{\sigma\in S_4/\Z_2^3}
\Tr(T^{a_{\sigma(1)}}T^{a_{\sigma(2)}})
\Tr(T^{a_{\sigma(3)}}T^{a_{\sigma(4)}})
A_{4;3}(\sigma(1)\sigma(2) \sigma(3)\sigma(4))
%\nonumber\\
\label{colordecomp}
\eea
We also define an explicit basis \cite{Glover:2001af} 
of single and double traces:
\bea
\label{basis}
&& \hspace{-5mm}
   \cC_\ii{1} = \Tr(T^{a_1} T^{a_2} T^{a_3} T^{a_4})\,, \qquad
   \cC_\ii{4} = \Tr(T^{a_1} T^{a_3} T^{a_2} T^{a_4})\,, \qquad
   \cC_\ii{7} = \Tr(T^{a_1} T^{a_2}) \Tr(T^{a_3} T^{a_4}) \nonumber\\
&& \hspace{-5mm}
   \cC_\ii{2} = \Tr(T^{a_1} T^{a_2} T^{a_4} T^{a_3})\,, \qquad
   \cC_\ii{5} = \Tr(T^{a_1} T^{a_3} T^{a_4} T^{a_2})\,, \qquad
   \cC_\ii{8} = \Tr(T^{a_1} T^{a_3}) \Tr(T^{a_2} T^{a_4}) \nonumber\\
&& \hspace{-5mm}
   \cC_\ii{3} = \Tr(T^{a_1} T^{a_4} T^{a_2} T^{a_3})\,, \qquad
   \cC_\ii{6} = \Tr(T^{a_1} T^{a_4} T^{a_3} T^{a_2})\,, \qquad
   \cC_\ii{9} = \Tr(T^{a_1} T^{a_4}) \Tr(T^{a_2} T^{a_3})  \nonumber\\
\eea
in terms of which the four-gluon amplitude can be expanded as
\be
\cA_{4} (1234)
=  g^2 \sum_{i=1}^{9} A_\ii{i}  \,\, \cC_\ii{i}
 \,.
\label{fourgluon}
\ee

The loop expansion of color-ordered amplitudes 
\be
A_\ii{i}  = \sumL a^L A^\Ell_\ii{i}  \, ,\;\;\;\;\; NA_{4;1}=\sumL a^L A_{4;1}^{(L)}\, ,\;\;\;\;
A_{4;3}=\sumL a^L A_{4;3}^{(L)}
\ee
is in terms of the natural 't Hooft loop expansion parameter \cite{Bern:2005iz}
\be
a \equiv {g^2 N \over 8 \pi^2} \left( 4 \pi \e^{-\gamma} \right)^\ep \,,
\ee
where $\gamma$ is Euler's constant,
and $\ep = (4-D)/2$. Note that at $L$ loops, the amplitude is at most of order $N^L$, which means that $A^\Ell_\ii{i}$ starts at ${\cal O}(N^0)$.

For a general $n$-point amplitude, we will have an expansion in an arbitrary number of multi-trace color ordered amplitudes 
$A_{n;j_1,j_2,...,j_k}$. 

Besides the loop expansion in the 't Hooft parameter $a$, we still have a $1/N$ expansion of the amplitudes, which can be understood in 't Hooft's 
double line notation as an expansion in the topology of the diagrams. For ${\cal A}_{4}$, the expansion in single-trace $A_{4;1}$ and 
double-trace $A_{4;3}$ amplitudes corresponds to the topology of the outside lines, forming boundaries of the diagrams. For example, at one-loop, the contribution in 
$A_{4;1}$ to the amplitude is leading, i.e. of order $N$ (thus $A_{4;1}$ of order 1), coming from a diagram with the topology of 4 external lines and a boundary, whereas the contribution 
of $A_{4;3}$ is subleading, i.e. of order $N^0$, and comes from a nonplanar diagram with 4 external lines, but arranged on two boundaries. 
It can be obtained 
by taking two twists of the 't Hooft double lines on opposite sides of the box, or twists on all 4 sides. Thus the multi-trace 
expansion comes as an expansion in the topology associated with the external lines (number of boundaries for them), 
and is an expansion in integer powers of $1/N$, 
corresponding to the number of boundaries of the diagram.

On top of that, we also have an expansion in integer powers of $1/N^2$, independently for $A_{4;1}$ and $A_{4;3}$, corresponding to nonplanar 
diagrams with handles (a handle gives a factor of $1/N^2$). The expansion terminates at order ${\cal O}(N^0)$ for the amplitude, since in the 
amplitude the powers of $N$ can only be positive. Thus at $L$-loops, we have 
$A_{4;1}^{(L)}={\cal O}(1)$ to ${\cal O}(1/N^{L})$
and $A_{4;3}^{(L)}={\cal O}(1/N)$ to ${\cal O}(1/N^{L})$.
Taken together, we will say that the gluon amplitudes have a $1/N$ expansion.

\section{IR divergences for subleading $\cN=4$ four-gluon amplitudes}

\subsection{General formalism}

$\cN=4$ SYM is a UV-finite theory,
but IR divergences arise due to the exchange
of soft and collinear gluons.
These divergences can be regulated using
dimensional regularization in  $D=4 - 2 \ep$ dimensions,
in which they appear as poles in a Laurent expansion in $\ep$.

In gluon-gluon scattering in ${\cN}=4$ SYM, 
IR divergences arise both from soft gluons 
and from collinear gluons,
each of which gives rise to an $\cO(1/\ep)$ pole at one loop, 
leading to an $\cO(1/\ep^2)$ pole at that order. 
At $L$ loops, the leading IR divergence of the 
scattering amplitude is therefore $\cO(1/\ep^{2L})$, 
arising from multiple soft gluon exchanges. 

Subleading-color amplitudes $A^\Ellk$,
that is, those suppressed by $1/N^k$ relative to the leading-color
amplitude at $L$ loops, have less severe IR divergences, 
being only of $\cO(1/\ep^{2L-k})$ at $L$-loops.

In this section,
we review the derivation of a compact all-loop-orders expression 
for the IR-divergent part 
of the $\cN=4$ SYM four-gluon amplitude 
given in ref.~\cite{Naculich:2008ys,Naculich:2009cv}.
This result is expressed in terms of the soft (cusp) anomalous dimension 
$\gamma(a)$,
the collinear anomalous dimension 
$\cG_0(a)$,
and the soft anomalous dimension matrices $\bGam^\pel$, 
and relies on the assumption that the soft anomalous dimension 
matrices are mutually commuting, 
which follows if they are all proportional to $\bGam^\One$,
as has been conjectured in 
refs.~\cite{Bern:2008pv,Becher:2009cu,Gardi:2009qi,Becher:2009qa}. 
This compact expression is then used 
to obtain the coefficient of the leading IR pole (and some
subleading poles) of all the subleading-color amplitudes.
Explicit values for the anomalous dimensions can be obtained
by comparison with various exact results.

We organize the 4-point color-ordered amplitudes
$A_\ii{i}$ defined in \eqn{fourgluon}
into a vector in color space
\cite{Catani:1996jh,%Catani:1996vz,
Catani:1998bh} 
\be
\ket{A}=\left(
A_\ii{1},  \,
A_\ii{2}, \,
A_\ii{3}, \,
A_\ii{4},  \,
A_\ii{5},  \,
A_\ii{6},  \,
A_\ii{7}, \, 
A_\ii{8}, \, 
A_\ii{9}  \right)^T
\ee
where $( \cdots )^T$ denotes the transposed vector.
The vector of color-ordered amplitudes factorizes into 
\cite{Sterman:2002qn,MertAybat:2006mz}
\be
\label{factorize}
\left|   A \left(\mommu,  a, \ep\right) \right> 
= 
 J \left(\Qmu, a, \eps \right) \, 
{\bS} \left( \mom,\Qmu,  a, \eps\right) 
\left | H \left( \mom,\Qmu,  a, \eps \right) \right>
\ee
where $\ket{H}$, which is IR-finite as $\eps \to 0$,
characterizes the short-distance behavior of the amplitude,  
and where the prefactors $J$ and ${\bS}$ 
encapsulate the long-distance IR-divergent behavior.
The soft function ${\bS}$
is written in boldface to denote that it 
is a matrix acting on the vector $\ket{H}$.
Also $s_{ij}$ is the kinematic invariant $ (k_i + k_j)^2$, 
$\mu$ is a renormalization scale,
and $Q$ is an arbitrary factorization scale
which serves to separate the long- and short-distance behavior. 

Because $\cN=4$ SYM theory is conformally invariant,
the product of jet functions $J$ 
may be explicitly evaluated as \cite{Bern:2005iz}
\be
J \left( \Qmu, a, \eps \right) 
= \exp \left[ -\frac{1}{2} \sum_{\ell=1}^\infty a^\ell
\muQell
\left( { \gamma^\pel \over  (\ell \eps)^2 }
        + { 2 \cG_0^\pel \over  \ell \eps }
\right)
\right]
\ee
where 
$\gamma^\pel$ and $\cG_0^\pel$ are the coefficients of 
the soft (or Wilson line cusp) and collinear anomalous dimensions 
of the gluon respectively.
The explicit values for these anomalous dimensions 
may be obtained from the exact expressions for the planar four-gluon amplitude
\cite{Bern:2005iz}:
\bea
\label{defanom}
\gamma (a)
&=& 
\sum_{\ell=1}^\infty a^\ell  \gamma^\pel = 
4 a - 4 \zeta_2 a^2 + 22 \zeta_4 a^3 + \cdots
\nonumber\\
\cG_0(a)
&=& 
\sum_{\ell=1}^\infty  a^\ell  \cG_0^\pel =
    -  \zeta_3 a^2 + (4 \zeta_5 + \frac{10}{3} \zeta_2 \zeta_3 )  a^3 + \cdots
\eea
The soft function ${\bS}$
is given by \cite{Sterman:2002qn,MertAybat:2006mz}
\be
{\bS} \left( \mom, \Qmu, a, \eps \right) 
\,=\,
{\rm P}~{\rm exp}\left[
\, -\; \frac{1}{2}\int_{0}^{Q^2} \frac{d\tilde{\mu}^2}{\tilde{\mu}^2}
\bGam \left( \mom,
             {\bar a}  \left(\frac{\mu^2}{\tilde{\mu}^2}, a, \eps  \right)
       \right) 
\right]\,
\label{soft}
\ee
where
\be
\bGam \left( \mom, a \right)
= \suml a^\ell \bGam^\pel, \qquad\qquad
{\bar a}  \left(   \frac{\mu^2}{\tilde{\mu}^2}, a, \eps \right) 
=  \left(     \frac{\mu^2}{\tilde{\mu}^2}   \right)^\eps  a.
\ee
suppressing the explicit dependence of $\bGam^\pel$ 
on ${ s_{ij} / Q^2 }$ to lighten the notation.

At this point, we make the assumption 
that the soft anomalous dimension matrices $\bGam^\pel$
all commute with one 
another\footnote{\label{gammafootnote}This 
assumption was also used to simplify the
IR divergences of QCD in ref.~\cite{Becher:2009qa}.
The assumption is certainly valid through two loops,
since $ \bGam^\Two = {1 \over 4} \gamma^\Two \bGam^\One $,
as shown in ref.~\cite{MertAybat:2006wq,MertAybat:2006mz}.
In ref.~\cite{Dixon:2009gx}, it was established that 
$ \bGam^\Three = {1 \over 4} \gamma^\Three \bGam^\One$
for the non pure gluon contributions.
Further, 
$ \bGam^\Ell = {1 \over 4} \gamma^\Ell \bGam^\One $
has been conjectured to hold to all orders 
in refs.~\cite{Bern:2008pv,Becher:2009cu,Gardi:2009qi,Becher:2009qa}.
Difficulties may arise at four loops, however,
due to the possibility of quartic Casimir 
terms \cite{Gardi:2009qi,Dixon:2009gx,Armoni:2006ux,Alday:2007mf}.}
so that the path ordering in \eqn{soft}
becomes irrelevant, 
allowing us to explicitly integrate it, obtaining 
\be
{\bS} \left( \mom, \Qmu, a, \eps \right) 
 = \exp \left[ \frac{1}{2} \sum_{\ell=1}^\infty a^\ell
\muQell
{ \bGam^\pel   \over  \ell \eps }
\right].
\ee
Combining the exponents of the jet and soft functions 
into \cite{Sterman:2002qn,Naculich:2008ys}
\be
\bG^\pel (\ep) =
\frac{N^\ell}{2}
\muQone
\left[-\left(
\frac{\gamma^\pel}{\ep^2}
+\frac{2\cG_0^\pel} {\ep}  \right)\bone 
+\frac{1}{\ep}\bGam^\pel\right]
\label{defG}
\ee
we may express the four-gluon amplitude 
in the compact form\footnote{Henceforth we suppress 
$ s_{ij} $, $Q$, $\mu$, and $a$
in the arguments of the amplitudes.}
\be
\label{compact}
\ket{A (\eps) } 
=  \exp\left[ \suml {a^\ell \over N^\ell} \bG^\pel (\ell \eps) \right] 
\ket{H (\eps)} 
\ee
or equivalently
\be
\label{deffinite}
\ket{ H (\eps)} 
= \sumL a^L \ket{H^\Ellf(\eps)} 
= \left( \bone - \suml \frac{a^\ell }{N^\ell}\bF^\pel (\ep)   \right) 
\ket{A (\eps) }.
\ee
where the matrices  $\bF^\pel (\ep)$ will be defined below.
Expanding \eqn{deffinite} through three loops, 
we obtain the expressions given in refs.~\cite{Sterman:2002qn,Naculich:2008ys}
\bea
\ket{A^\One (\ep)} 
&=& \frac{1}{N} \bF^\One(\ep) \ket{A^\Zero} + \ket{H^\Onef (\ep)}
\nonumber
\\
\ket{A^\Two (\ep)} 
&=& \frac{1}{N^2} \bF^\Two(\ep) \ket{A^\Zero}  
+ \frac{1}{N} \bF^\One(\ep)\ket{A^\One (\ep)} 
+ \ket{H^\Twof (\ep)}  
\label{twoloopSTY}
\\
\ket{A^\Three (\ep)} 
&=& \frac{1}{N^3} \bF^\Three(\ep) \ket{A^\Zero}  
 +  \frac{1}{N^2} \bF^\Two(\ep) \ket{A^\One}  
+ \frac{1}{N} \bF^\One(\ep)\ket{A^\Two (\ep)} 
+ \ket{H^\Threef (\ep)}  
\nonumber
\eea
which will be useful in extracting the IR-divergent terms  
of leading- and subleading-color amplitudes in the following section.
(Note that, because of the presence of poles in $\bF(\eps)$,
we will need to know positive powers of $\ep$ in the expansion 
of lower loop amplitudes
to obtain all the  IR-divergent contributions to the $L$-loop
amplitude $A^\pel$.)

The equivalence of \eqns{compact}{deffinite}
follows if the matrices  $\bF^\pel (\ep)$ are defined through the equation
\be
\label{defF}
\left( \bone - \suml \frac{a^\ell }{N^\ell} \bF^\pel (\ep)  \right) 
\exp\left[ \suml \frac{a^\ell }{N^\ell}\bG^\pel (\ell \eps)   \right] 
= \bone \,.
\ee
First define the functional $X[M]$ via \cite{Bern:2005iz}
\be
\label{defX}
1 + \suml a^\ell M^\pel 
\equiv \exp \left[ \suml a^\ell \left( M^\pel - X^\pel[M] \right)\right]
\ee
so that $X^\One[M] = 0 $,
$X^\Two[M] = {1\over 2} \left[M^\One \right]^2 $,
$X^\Three[M] = -{1\over 3} \left[M^\One \right]^3
+ M^\One M^\Two $, etc.
This functional was defined for scalar functions $M^\pel$,   
but we can also use it for commuting matrices.
We have assumed that $\bGam^\pel$ and therefore $\bG^\pel$ 
are mutually commuting, 
and thus so are $\bF^\pel$,  as a result of \eqn{defF}.
Thus 
\be
\left( \bone - \suml \frac{a^\ell}{N^\ell}  \bF^\pel (\ep)  \right) 
= \exp \left[ \sumL \frac{a^\ell}{N^\ell}  
\left(- \bF^\pel (\eps) - X^\pel[-\bF] \right)\right]
\ee
and so \eqn{defF} is equivalent to 
\be
\bF^\pel  (\eps) = -  X^\pel[-\bF]  + \bG^\pel (\ell \eps) 
\label{recurse}
\ee
which defines $\bF^\pel$ recursively in terms of 
$\bG^\pel$ and $\bF^{(\ell')}$ with $\ell' < \ell$.
The explicit expressions up through three loops 
\bea
\bF^{\One}(\ep)
&=&  \bG^\One (\ep)
\nonumber\\
\bF^{\Two}(\ep)
&=&
- \half \left[ \bF^\One (\ep) \right]^2 + \bG^\Two (2 \ep)
\label{Ftwodef}\\
\bF^{\Three}(\ep)
&=&
- \third \left[ \bF^\One (\ep) \right]^3 
- \bF^\One (\ep) \bF^\Two (\ep)
+ \bG^\Three (3 \ep)
\nonumber
\eea
agree (up to rescaling by a factor of $N^L$) 
with the expressions 
given in ref.~\cite{Sterman:2002qn} 
when specialized to the case of $gg \to gg$ in $\cN=4$ SYM theory.

\subsection{$1/N$ expansion of IR divergences}

In this subsection, we will use the results of the previous subsection
to expand the IR-divergent contributions of the four-gluon
amplitude in powers of $1/N$.

The $L$-loop color-ordered amplitudes 
may be written in a $1/N$ expansion as
\be
\ket{A^\Ell(\eps)} = \sum_{k=0}^L \frac{1}{N^k} \ket{A^\Ellk(\eps)}
\label{oneoverN}
\ee
where $\ket{A^\Ellzero}$ are the leading-color amplitudes,
arising from planar diagrams,
and $\ket{A^\Ellk}$,  $1 \le k \le L$,
are the subleading-color amplitudes,
which include contributions from nonplanar diagrams as well.
The single-trace amplitudes 
($i=1, \ldots, 6$)
only contain even powers of $1/N$
(relative to the leading-color amplitude), 
while the double-trace amplitudes 
($i=7, \ldots, 9$) only contain odd powers of $1/N$.

We begin by expanding \eqn{compact}:
\be
\hspace{-1mm}
\ket{A(\eps)} = 
\sumL \sum_{k=0}^L  \frac{a^L}{N^k} \ket{A^\Ellk(\eps)} = 
\prod_{\ell=1}^\infty 
\sum_{\{n_\ell\}}  
{1 \over n_\ell !}
\left( a^\ell \frac{\bG^\pel (\ell \eps) }{N^\ell}  \right)^{n_\ell} 
\sum_{\ell_0=0}^\infty \sum_{k_0 = 0}^{\ell_0} {a^{\ell_0} \over N^{k_0}} 
\ket{H^{(\ell_0, k_0)}(\eps)}.
\label{expandinit}
\ee
In the derivation of \eqn{expandinit},
we assumed that the soft-anomalous dimension matrices are
mutually commuting.
We now assume further that the 
higher-loop soft-anomalous dimension matrices 
are all proportional to the one-loop soft-anomalous 
dimension matrix 
\be 
\bGam^\pel = {\gamma^\pel\over 4}  \bGam^\One
\qquad {\it (assumption)}
\label{prop}
\ee
as has been conjectured (see footnote \ref{gammafootnote}).
This allows us to rewrite \eqn{defG} as
\be
\label{newG}
\frac{\bG^\pel (\ell \ep)} {N^\ell}
 = 
  {1 \over 2}
\muQell
\left[-\left(
\frac{\gamma^\pel}{(\ell \ep)^2}
+\frac{2\cG_0^\pel} {\ell \ep}  \right)\bone 
+\frac{\gamma^\pel}{4 \ell \ep}\bGam^\One\right].
\ee
The one-loop soft anomalous dimension matrix can be written \cite{MertAybat:2006mz}
\be
\label{oneloopanom}
\bGam^\One
=
- \frac{1}{N}  \sum_{i=1}^4 \sum_{j\neq i}^4 \bT_i \cdot \bT_j
 \log \left( {-s_{ij} \over  Q^2 } \right) 
\ee
where $\bT_i \cdot \bT_j = T_i^a T_j^a$
with $T_i^a$ the SU$(N)$ generators in the adjoint representation.
In the basis (\ref{basis}), it has the explicit form \cite{Glover:2001af}
\be
\label{defGamOne}
\bGam^\One = 
2 \left( 
\begin{array}{cc}
 \ia &     0 \\
  0  &  \id 
\end{array}
\right)
+ {2 \over N} \left( \begin{array}{cc} 0 & \ib \\
\ic  & 0
\end{array}
\right)
\ee 
where 
\bea
\label{defalpha}
\ia  = 
\left( {
\begin{array}{cccccc}
\tS+\tT & 0 & 0 & 0 & 0 & 0 \\
0 & \tS+\tU & 0 & 0 & 0 & 0 \\
0 & 0 & \tT+\tU & 0 & 0 & 0 \\
0 & 0 & 0 & \tT+\tU & 0 & 0\\
0 & 0 & 0 & 0 & \tS+\tU & 0\\
0 & 0 & 0 & 0 & 0 & \tS+\tT
\end{array} } \right), &&
\ib = 
\left( {
\begin{array}{ccc}
\tT-\tU & 0 & \tS-\tU \\
\tU-\tT & \tS-\tT & 0 \\
0 & \tT-\tS & \tU-\tS \\
0 & \tT-\tS & \tU-\tS \\
\tU-\tT & \tS-\tT & 0 \\
\tT-\tU & 0 & \tS-\tU \\
\end{array}
}\right)
\nonumber\\[4mm]
\ic  = 
\left( {
\begin{array}{cccccc}
\tS-\tU & \tS-\tT & 0 & 0 & \tS-\tT & \tS-\tU \\
0 & \tU-\tT & \tU-\tS & \tU-\tS & \tU-\tT& 0 \\
\tT-\tU & 0 & \tT-\tS & \tT-\tS & 0 & \tT-\tU \\
\end{array}
}\right), &&
\id = 
\left( {
\begin{array}{ccc}
2\tS & 0 & 0 \\
0 & 2\tU & 0 \\
0 & 0 & 2\tT
\end{array}
} \right)
\eea
with
\be
\tS = \log \left(-\frac{s}{Q^2}\right),\qquad\qquad 
\tT = \log \left(-\frac{t}{Q^2}\right), \qquad\qquad 
\tU = \log \left(-\frac{u}{Q^2}\right)\,.
\ee
If the assumption (\ref{prop}) is valid, 
then the $1/N$ expansion of ${\bG^\pel (\ell \eps) }/{N^\ell} $
terminates after two terms
\be
\frac{\bG^\pel (\ell \eps) }{N^\ell}   
=
\gzell  + {1 \over N} \goell
\ee
where $\gzell$ and $\goell$ can be read from \eqns{newG}{defGamOne}.
We rewrite \eqn{expandinit}  as
\be
\hspace{-1mm} 
\ket{A(\eps)} = 
\sumL \sum_{k=0}^L  \frac{a^L}{N^k} \ket{A^\Ellk(\eps)} = 
\prod_{\ell=1}^\infty 
\sum_{\{n_\ell\}}  
{1 \over n_\ell !}
\left( a^\ell \gzell + {a^\ell \over N} \goell  \right)^{n_\ell} 
\sum_{\ell_0=0}^\infty \sum_{k_0 = 0}^{\ell_0} {a^{\ell_0} \over N^{k_0}} 
\ket{H^{(\ell_0, k_0)}(\eps)}
\label{expand}
\ee
making all $N$ dependence explicit.

We now determine the power of the leading IR pole of
$\ket{A^\Ellk (\eps)}$.
Consider an individual term on the r.h.s.~of \eqn{expand}.
By counting powers of $a$ and $1/N$, one sees that 
this term contributes to $\ket{A^\Ellk(\eps)}$, with
\be
\label{defLk}
L= \ell_0 + \suml   \ell n_\ell  , \qquad
k=k_0+k_1
\ee
where $k_1$ is the number of factors $\goell$ present in the term.
{}From \eqns{newG}{defGamOne},
it is apparent that $\gzell$ has a double pole in $\eps$,
but $\goell$ only has a single pole. 
The leading IR pole in the term under consideration is 
therefore $1/\eps^p$, where
\be
p =2 \suml n_\ell - k_1\,.
\label{firstp}
\ee
Combining \eqns{defLk}{firstp}, 
we find
\be
p = 2 L - k 
- \left[ 2 \suml (\ell -1) n_\ell  + 2 \ell_0 - k_0 \right].
\label{defp}
\ee
Since $k_0 \le \ell_0$, 
the term in square brackets is non-negative, and we conclude that 
\be
\ket{A^\Ellk(\eps)}  \sim \cO \left( \frac{1}{\eps^{2L-k}}\right).
\ee
This behavior was previously conjectured in ref.~\cite{Naculich:2008ys} 
and shown in ref.~\cite{Naculich:2009cv} 
(subject to the assumptions stated above). 

Next we review the derivation \cite{Naculich:2008ys,Naculich:2009cv}  
of the coefficient of the
leading IR pole of $\ket{A^\Ellk(\eps)}$.
Terms in \eqn{expand} contribute to the leading
IR pole only when the expression in square brackets in \eqn{defp}
vanishes,
which occurs when $n_\ell=0$ for $\ell \ge 2$, and $\ell_0=k_0=0$
(with $n_1$ unconstrained).     
In other words, the leading IR divergences are given
 by \cite{Naculich:2008ys,Naculich:2009cv}
\be
\ket{A(\eps)}  
\sim  \exp\left[  a \frac{\bG^\One (\eps) }{N}  \right] \ket{\Azero}
\qquad {\it (leading~IR~divergence)}
\label{leading}
\ee

Recalling that
\be
\frac{\bG^\One (\eps) }{N}   
= \muQone
\left[ - {2 \over \eps^2} \bone 
    + {1 \over \eps}
\left( \begin{array}{cc}
\ia &   0   \\
  0    & \id 
\end{array} \right)
+ {1 \over N \eps}
\left( \begin{array}{cc}
  0 & \ib   \\
  \ic   & 0
\end{array} \right)
\right]
\ee
we use \eqn{leading} to obtain 
the coefficient of the leading IR pole
\be
\ket{A^\Ellk (\eps)} = 
{(-2)^{L-k} \over k! (L-k)!}  
{1 \over \eps^{2L-k}}
\left( \begin{array}{cc}
  0 & \ib   \\
  \ic   & 0
\end{array} \right)^k \ket{\Azero}
+ \cO\left( 1 \over \eps^{2L-k-1} \right).
\ee
where the tree-level amplitudes are 
\be
\ket{H^{(0,0)}}  = \ket{\Azero} 
= 
- \frac{ 4 i K }{stu}
\left(u, t, s, s, t, u, 0, 0, 0\right)^T
\label{tree}
\ee 
where $s=(k_1+k_2)^2$, $t=(k_1+k_4)^2$,  and $u=(k_1+k_3)^2$
are the usual Mandelstam variables,
obeying $s+t+u=0$ for massless external gluons.
The factor $K$, defined in eq.~(7.4.42) of ref.~\cite{Green:1987sp},
depends on the momenta and helicity of the external gluons, 
and is totally symmetric under permutations of the external legs.

The leading IR pole of the planar amplitude is simply
\be
\ket{A^\Ellzero (\eps)} = 
{(-2)^{L} \over L! ~\eps^{2L}}  \ket{\Azero}
+ \cO\left( 1 \over \eps^{2L-1} \right) \,.
\ee
The remaining IR divergences,
from $\cO(1/\ep^{2L-1})$ to $\cO(1/\ep)$,
are all proportional to $\ket{\Azero} $
and are given by the (generalized) ABDK equation \cite{Bern:2005iz} 
(see appendix A of ref.~\cite{Naculich:2009cv} ).

We now write an explicit expression for the 
coefficients of the leading IR poles of 
{\it subleading-color} amplitudes.
First we use \eqns{tree}{defalpha} to show
\be
\label{gammabeta}
\ic 
\begin{pmatrix}u \cr t \cr s \cr s \cr t \cr u\end{pmatrix}
=
2(sY-tX) \ones,
\qquad\quad{\rm and}\quad\qquad
\gamma \beta  \ones
= 
2 \left( X^2 + Y^2 + Z^2 \right)
\ones
\ee
with
\be
\label{defXYZ}
X = \log \left(t \over u\right), \qquad
Y = \log \left(u \over s\right), \qquad
Z = \log \left(s \over t\right).
\ee
Hence, the leading IR divergence of the subleading-color amplitudes
is given by 
\be
\ket{A^\Ellodd (\eps)}
=
\left({-4 i K \over stu} \right)
{(-1)^{L-1} 2^{L-m}  
\left( X^2 + Y^2 + Z^2 \right)^m 
(sY-tX)
\over 
(2m+1)! (L-2m-1)!  
\,\eps^{2L-2m-1}}
\begin{pmatrix}
0\cr
0\cr
0\cr
0\cr
0\cr
0\cr
1\cr
1\cr
1\cr\end{pmatrix}
+ 
\cO\left( 1 \over \eps^{2L-2m-2} \right)
\label{odd}
\ee
and
\be
\ket{A^\Ellevenplustwo (\eps)}
=
\left({-4 i K \over stu} \right)
{
(-1)^{L} 2^{L-m-1}  
\left( X^2 + Y^2 + Z^2 \right)^m 
(sY-tX)
\over 
(2m+2)! (L-2m-2)!
\eps^{2L-2m-2}
}  
\begin{pmatrix}
X-Y\cr
Z-X\cr
Y-Z\cr
Y-Z\cr
Z-X\cr
X-Y\cr
0\cr
0\cr
0 \cr\end{pmatrix}
+
\cO\left( 1 \over \eps^{2L-2m-3} \right)
\label{even}
\ee
The results (\ref{odd}) and (\ref{even})
were derived in ref.~\cite{Naculich:2009cv}, 
generalizing expressions derived in ref.~\cite{Naculich:2008ys}.

\subsection{IR divergences of $A^\Ellone$}
\label{secIREllone} 

In this subsection, we consider the subleading-color amplitude 
$\ket{A^\Ellone}$,
and derive the first 
three\footnote{It is straightforward to obtain further terms 
in the Laurent expansion as needed.}
terms in the Laurent expansion.
Consider all terms in \eqn{expand}
for which the expression in square brackets in \eqn{defp}
is $\le 2$:
\bea
\ket{A^\Ell(\eps)} 
&=& \frac{1}{L!} 
    \left( \gz + {1\over N} \gone \right)^L \ket{\Azero}
+ \frac{1}{N (L-1)!} 
    \left( \gz + {1\over N} \gone       \right)^{L-1} 
						\ket{H^{(1,1)}(\eps)}
\label{firstthree}
\\
&+&
 \frac{1}{ (L-2)!} 
    \left( \gz + {1\over N} \gone       \right)^{L-2} 
    \left( \gztwo + {1\over N} \gotwo          \right) \ket{\Azero}
+ \frac{1}{ (L-1)!} 
    \left( \gz + {1\over N} \gone       \right)^{L-1} 
						\ket{H^{(1,0)}(\eps)}
\nonumber\\
&+&
 \frac{1}{ N^2 (L-2)!} 
    \left( \gz + {1\over N} \gone       \right)^{L-2} 
						\ket{H^{(2,2)}(\eps)}
+ \cdots
\qquad\qquad {\it (three~leading~IR~poles)}
\nonumber
\eea
where we use \eqns{newG}{defGamOne} to write 
\bea
\gz
= \muQone \left[- {2 \over \eps^2} \bone 
  + {1 \over \eps}
\left( \begin{array}{cc}
\ia &   0   \\
  0    & \id 
\end{array} \right) \right],
&&
\gone
=
{1 \over \eps}
\muQone
\left( 
\begin{array}{cc}
  0 & \ib   \\
  \ic   & 0
\end{array} \right),
\nonumber\\
\gztwo=
\muQtwo \left[
 - \left( {\gamma^\Two \over 8 \eps^2} + {\cG_0^\Two \over 2 \eps} \right)
 \bone 
+ {\gamma^\Two \over 8 \eps} 
\left( \begin{array}{cc}
\ia &   0   \\
  0    & \id 
\end{array} \right) \right],
&&
\gotwo
=
{\gamma^\Two \over 8 \eps} 
\muQtwo
\left( 
\begin{array}{cc}
  0 & \ib   \\
  \ic   & 0
\end{array} \right).
\qquad
\qquad
\label{onetwo}
\eea
To extract the $\ket{A^\Ellone}$ amplitude,
we employ the identity
\bea
&&\left( \gz + {1\over N} \gone \right)^L  \Bigg|_{1/N~{\rm piece}}
\label{ident}
\\
&& \quad=
L \gz^{L-1} \gone
~+~ {L \choose 2} \gz^{L-2} [\gone,\gz]
~+~ {L \choose 3} \gz^{L-3} [ [\gone,\gz], \gz]
%+ {L \choose 4} \gz^{L-4} [ [ [\gone,\gz], \gz], \gz] 
~+~  [ \cdots [[[\gone,\gz], \gz], \gz]  \cdots ]
\nonumber
\eea
in which the first term on the r.h.s. has an expansion 
that starts with $1/\eps^{2L-1}$,
the second term has an expansion 
that starts with $1/\eps^{2L-2}$,
and so forth.
Thus, keeping only the 
terms proportional to $1/N$ in \eqn{firstthree},
and only the first three terms in the Laurent expansion,
we obtain
\bea
\ket{A^\Ellone} 
&=& 
\frac{1}{(L-1)!} \gz^{L-1} \gone \ket{\Azero}
+ \frac{1}{2 (L-2)!} \gz^{L-2} [\gone,\gz] \ket{\Azero}
+ \frac{1}{(L-1)!} \gz^{L-1} \ket{H^{(1,1)}(\eps)} 
\nonumber\\
&+&
 \frac{1}{6(L-3)!} \gz^{L-3} [ [\gone,\gz], \gz] \ket{\Azero}
+ \frac{1}{ (L-2)!} 
    \gz^{L-2} \gotwo \ket{\Azero}
+ \frac{1}{ (L-3)!} 
    \gz^{L-3} \gone \gztwo  \ket{\Azero}
\nonumber\\
&+&
 \frac{1}{ (L-2)!}  \gz^{L-2} \gone \ket{H^{(1,0)}(\eps)}
+ \cO \left( \frac{1}{\eps^{2L-4}} \right) .
\label{firstsubleading}
\eea
as obtained in ref.~\cite{Naculich:2009cv}, 

\subsection{IR divergences of $A^\EllL$}

In this subsection, 
we derive an expression 
for the coefficient of the IR divergences of the 
first two terms in the Laurent expansion 
of the most subleading-color amplitude $\ket{A^\EllL}$.

The only terms in \eqn{expand} that contribute to $\ket{A^\EllL}$
are those with as many factors
of $1/N$ as of $a$. 
Thus, only $\gone$ and $ \ket{H^{(\ell_0, \ell_0)}}$ can contribute,
giving 
\be
\hspace{-1mm}
\ket{A^\EllL(\eps) } 
= \sum_{\ell_0=0}^L   {1\over (L-\ell_0)!} 
\gone^{L-\ell_0}
\ket{H^{(\ell_0,\ell_0)}(\eps)},
\qquad
{\rm where}
\quad
\gone
=
{1 \over \eps}
\muQone
\left( \begin{array}{cc}
  0 & \ib   \\
  \ic   & 0
\end{array} \right)
\ee
exact to all orders in the $\eps$ expansion.
Keeping just the first two terms in the Laurent expansion, we find
\bea
\ket{A^\EllL(\eps)}
 &=& 
\frac{1}{(L-1)!} 
\gone^{L-1}
\left[ \frac{1}{L} \gone \ket{\Azero} 
+ \ket{H^{(1,1)}(\eps)}  \right]
+ \cO\left( 1 \over \eps^{L-2} \right)
\nonumber
\\ 
&=&
\frac{1}{(L-1)!} 
{1\over \eps^{L-1}} 
\left( \begin{array}{cc}
  0 & \ib   \\
  \ic   & 0
\end{array} \right)^{L-1}
\ket{A^{(1,1)}(L \eps)}  + \cO\left( 1 \over \eps^{L-2} \right).
\label{mostsubleading}
\eea
This was derived in ref.~\cite{Naculich:2009cv}, 
and confirms the conjecture made in eqs.~(4.45) and (4.46) 
of ref.~\cite{Naculich:2008ys}.

\subsection{Exact expressions at one- and two-loops} 

$\cN=4$ SYM scattering amplitudes
may be expressed in terms of planar and nonplanar scalar loop integrals.
The two-loop four-gluon scattering amplitude was first computed 
by Bern, Rosowsky, and Yan \cite{Bern:1997nh}
(see also ref.~\cite{Bern:1998ug}).
Explicit expressions for these IR-divergent integrals
as Laurent expansions in $\ep$ 
were later obtained by Smirnov in the planar case \cite{Smirnov:1999gc},
and by Tausk in the non-planar case \cite{Tausk:1999vh}.
In this subsection, we review these results and some formulas for the $1/N$ expansion of these divergences.

Recall from \eqn{oneoverN} that
$A_\ii{i}^\Ellk$
denotes the $L$-loop color-ordered amplitude
which is subleading by a factor of $1/N^k$ in the $1/N$ expansion.
Single-trace amplitudes are denoted by $i=1, \ldots, 6$ 
and double-trace amplitudes by $i=7, \ldots, 9$ (see \eqn{basis}). 

At one loop, the single-trace amplitudes are given by \cite{Green:1982sw}
\be
\label{oneloopST}
A^\Onezero_\ii{1} 
=  M^\One (s,t) \, A^\Zero_\ii{1}
=  2 i K \, I_4^\One(s,t)
\ee
with the other 
single-trace amplitudes $A^\Onezero_\ii{2}$ and $A^\Onezero_\ii{3}$
obtained by letting $t \lr u$ and $s \lr u$ respectively. 
The identities 
$A^\Ell_\ii{1} = A^\Ell_\ii{6}$,
$A^\Ell_\ii{2} = A^\Ell_\ii{5}$,
and 
$A^\Ell_\ii{3} = A^\Ell_\ii{4}$
are satisfied at all loop orders. 
In eq.~(\ref{oneloopST}),
$I_4^\One (s,t)$ denotes the scalar box integral
\bea
M^\One (s,t) &=& - \half st\, I_4^\One (s,t)
\label{Monedef}
\\
I_4^\One (s,t)
= I_4^\One (t,s)
&=& -i \mu^{2\ep} \e^{\ep \gamma} \pi^{-D/2}  \int 
{d^D p \over p^2 (p-k_1)^2 (p-k_1-k_2)^2 (p+k_4)^2 } 
\nonumber
\eea
an explicit expression for which is given, e.g., in ref.~\cite{Bern:2005iz}.

The one-loop double-trace amplitudes are given by \cite{Green:1982sw}
\bea
\label{oneloopDT}
A^\Oneone_\ii{7} =
A^\Oneone_\ii{8} =
A^\Oneone_\ii{9} 
&=&
2 \left( A^\Onezero_\ii{1} + A^\Onezero_\ii{2} + A^\Onezero_\ii{3}  \right)
\label{oneloopdecouple}
\\
&=&
4 i K \left[ I_4^\One (s,t)+ I_4^\One (t,u)+ I_4^\One (u,s) \right] \,.
\label{oneloopnp}
\eea
The relation (\ref{oneloopdecouple}) follows from the 
one-loop U(1) decoupling identity \cite{Bern:1990ux}.

At two loops, the leading-color single-trace amplitude 
is given by  \cite{Bern:1997nh}
\be
\label{twoloopLC}
A_\ii{1}^\Twozero 
= M^\Two (s,t) \, A^\Zero_\ii{1}
= - i K \left[ s I_4^\TwoP (s,t) +t I_4^\TwoP (t,s) \right]
\ee
where $I_4^\TwoP(s,t)$ denotes the scalar double-box (planar) integral 
\bea
M^\Two(s,t) &=&
  \frac{1}{4} st\, \left[ s I_4^\TwoP (s,t) +t I_4^\TwoP (t,s) \right]
\label{Mtwodef}
\\
 I_4^\TwoP(s,t) &=&
\left( -i \mu^{2\ep} \e^{\ep \gamma} \pi^{-D/2} \right)^2
\int 
 {d^D p \, d^D q
\over p^2 \, (p + q)^2 q^2 \, (p - k_1)^2 \,(p - k_1 - k_2)^2 \,
        (q-k_4)^2 \, (q - k_3 - k_4)^2 } 
\nonumber
\eea
an explicit expression for which is given, e.g., in ref.~\cite{Bern:2005iz}.
The double-trace amplitude is \cite{Bern:1997nh}
\bea
\hspace{-5mm}
A_\ii{7}^\Twoone &=& 
-2 i K
\Bigl[
s \left( 3 I_4^\TwoP(s, t) + 2 I_4^\TwoNP(s, t)
+ 3 I_4^\TwoP(s, u)  + 2 I_4^\TwoNP(s, u)\right)
\label{twoloopDT}
\\ && \hspace{11mm}
-t \left(I_4^\TwoNP(t, s) + I_4^\TwoNP(t, u)\right)
-u \left(I_4^\TwoNP(u, s) + I_4^\TwoNP(u,t)\right)
\Bigr]  
\nonumber
\eea
and the subleading-color single-trace amplitude is \cite{Bern:1997nh}
\bea
\hspace{-9mm}
A_\ii{1}^\Twotwo &=& 
-2 i K \,
\Bigl[
s \left(I_4^\TwoP(s, t) + I_4^\TwoNP(s, t)
+ I_4^\TwoP(s, u)  + I_4^\TwoNP(s, u)\right)
\label{twoloopSC}
\\ && \hspace{11mm}
+t \left(I_4^\TwoP(t, s) + I_4^\TwoNP(t, s)
+ I_4^\TwoP(t, u)  + I_4^\TwoNP(t, u)\right)
\nonumber\\ && \hspace{10mm}
-2u \left(I_4^\TwoP(u, s) + I_4^\TwoNP(u, s)
+ I_4^\TwoP(u,t)  + I_4^\TwoNP(u,t)\right)
\Bigr]  
\nonumber
\eea
where $I_4^\TwoNP(s,t)$ denotes the two-loop non-planar integral
\begin{equation} 
I_4^\TwoNP(s,t)  = 
\left( -i \mu^{2\ep} \e^{\ep \gamma} \pi^{-D/2} \right)^2 \int 
{d^D p \, d^D q\over p^2\,(p+q)^2\, q^2 \, (p-k_2)^2  \,(p+q+k_1)^2\,
  (q-k_3)^2 \, (q-k_3-k_4)^2}  
\label{twoloopNP}
\ee
an explicit expression for which is given in ref.~\cite{Tausk:1999vh}.
All the other single- and double-trace amplitudes $A^\Twok_\ii{i}$ 
are obtained by making the appropriate permutations of $s$, $t$, and $u$ 
in these expressions.

It is well-known \cite{Bern:2005iz} 
that planar amplitudes have the property of uniform transcendentality.
It is less obvious but nevertheless true \cite{Naculich:2008ys}
that subleading-color $\cN=4$ amplitudes 
at one and two loops (and presumably beyond) 
also have uniform transcendentality.  
What makes this surprising
is that the non-planar integral $I_4^\TwoNP(s,t)$ that contributes to 
$A^\Twoone$ and $A^\Twotwo$ does not have uniform transcendentality \cite{Naculich:2008ew,Brandhuber:2008tf}.
The subleading transcendentality parts, however, cancel out in the expressions 
(\ref{twoloopDT}) and (\ref{twoloopSC}).
(The same thing happens for the two-loop four-point amplitude of $\cN=8$ supergravity \cite{Naculich:2008ew,Brandhuber:2008tf}.)

The two-loop amplitudes obey the following group theory 
relations \cite{Bern:2002tk}
\bea
A_\ii{7}^\Twoone &=& 
2 \left( A^\Twozero_\ii{1} + A^\Twozero_\ii{2} + A^\Twozero_\ii{3}  \right)
- A_\ii{3}^\Twotwo \nonumber\\
A_\ii{8}^\Twoone &=& 
2 \left( A^\Twozero_\ii{1} + A^\Twozero_\ii{2} + A^\Twozero_\ii{3}  \right)
- A_\ii{1}^\Twotwo \nonumber\\
A_\ii{9}^\Twoone &=& 
2 \left( A^\Twozero_\ii{1} + A^\Twozero_\ii{2} + A^\Twozero_\ii{3}  \right)
- A_\ii{2}^\Twotwo
\label{STDT}
\eea
and may be easily verified using 
eqs.~(\ref{twoloopLC}), 
(\ref{twoloopDT}), 
and (\ref{twoloopSC}).
In addition, we have
\be
\label{sumSCvanishes}
A^\Twotwo_\ii{1} + A^\Twotwo_\ii{2} + A^\Twotwo_\ii{3} =0
\ee
also easily verified using \eqn{twoloopSC}.
Together these equations imply
\be
6 \, \sum_{i=1}^3   A_\ii{i}^\Twozero -
\sum_{i=7}^9   A_\ii{i}^\Twoone =0
\label{twoloopdecouple}
\ee
which is the two-loop generalization of the U(1) decoupling 
relation (\ref{oneloopdecouple}).
Both \eqns{sumSCvanishes}{twoloopdecouple} are encapsulated in the
equation
\be
6\, \sum_{i=1}^3   A_\ii{i}^\Ell -
N \,\sum_{i=7}^9   A_\ii{i}^\Ell =0, \qquad L \le 2
\ee
which is 
valid through two loops.

At one-loop, we also saw that one can relate all the subleading-color amplitudes $A_{n;j}$ to the leading amplitude $A_{n;1}$ via the group theory 
relation (\ref{subleadingrel}).

We now list some explicit formulas for the IR-divergent pieces
of one and two-loop amplitudes that will be of use in the following section.
At one-loop, the leading 4-point amplitude is 
given by \eqn{oneloopST} with 
\bea
M^\One (s,t) &=& \, - \, \frac{1}{\ep^2}   \left(\mu^2 \over -s\right)^\ep
                 \, - \, \frac{1}{\ep^2}   \left(\mu^2 \over -t\right)^\ep
+ \half \log^2 \left( s\over t\right) + {2 \pi^2\over 3}
+ \cO(\ep) 
\label{Mone}
\eea
while the exact relation (\ref{oneloopDT})
can be used to write both the IR-divergent and IR-finite
contributions to the double-trace subleading-color amplitude
\be
\ket{ A^\OneDT (\ep)} 
=
\Atree \left[ 
\left( \mu^2 \over -u \right)^\ep
\frac{(s Y - t X)}{\ep}  
- (s+t) XY 
 \right]  \ones
~+~\cO(\ep)
\label{oneDTexpand}
\ee
where we have only included the $[7]$, $[8]$, and $[9]$ components
of $A_\ii{i}^\OneDT$ as the others vanish.

At two loops, the planar amplitude is given by \eqn{twoloopLC} 
with \cite{Anastasiou:2003kj} 
\bea
M^\Two (\ep) 
&=&
\half \left[ M^\One (\ep)\right]^2 
- (\zeta_2  + \ep \zeta_3 + \ep^2 \zeta_4)  M^\One (2\ep)  - \frac{\pi^4}{72}
+ \cO(\ep)\,.
\label{ABDK}
\eea
The two-loop double trace amplitude has an IR divergence given by the general formula (\ref{odd}), which yields
\be
\ket{A^\TwoDT (\ep)} =
\Atree 
\frac{(-2)(s Y - t X)}{\ep^3}  
 \ones
~+~ \cO\left(1 \over \ep^2\right)
\ee
Finally, the subleading-color single-trace amplitude is given by 
\eqn{mostsubleading} which in this case yields
\be
\ket{A^\TwoSC (\ep)} 
=
\frac{1}{\ep}
\begin{pmatrix}X-Y \cr
         Z-X\cr 
         Y-Z\cr
         Y-Z\cr
         Z-X\cr 
	X-Y \cr\end{pmatrix}
A^{\OneDT}_{[7]}(2\ep) +
{\cal O}(\ep^0) \,.
\label{twoloopmat}
\ee
Only the [1] through [6] components are listed, as the
[7] through [9] components vanish.

%We note that at one and two-loops, the IR divergent Catani operator operator $I^{(l)}(\epsilon)$, related to the 
%$F^{(l)}(\epsilon)$ (including subleading pieces) has maximal uniform transcendentality, while at 3-loops the same is true
%\cite{Naculich:2008ys}. We expect then that 
%the IR divergent Catani operator $I^{(l)}(\epsilon)$ continues to have maximal uniform transcendentality, even at higher loops.
%Note also that the supergravity amplitudes have again maximal uniform transcendentality up to two loops \cite{Naculich:2008ew}. 
%The planar integrals (\ref{Monedef}) and (\ref{Mtwodef}) satisfy it, but the nonplanar integral (\ref{twoloopNP}) doesn't, and 
%only after taking the permutations needed for the ${\cal N}=8$ supergravity amplitude do we get a result of maximal 
%uniform transcendentality.

\section{Subleading color amplitudes of ${\cal N}=4$ SYM and amplitudes of ${\cal N}=8$ supergravity}

The $AdS_5/CFT_4$ correspondence provides a strong/weak duality between ${\cal N}=4$ SYM and ${\cal N}=8$ supergravity. These relationships have 
been extensively explored and exploited. There are also numerous indications of a weak/weak duality between the two theories, although this latter 
possibility is much less developed. Nevertheless this may be a very fruitful approach in attempts to understand relationships between the two 
theories. 
A lot of work has been done to relate the perturbation expansions of these two theories \cite{Naculich:2008ys,Bern:1998ug,Green:1982sw,Berends:1988zp,BjerrumBohr:2010yc,BjerrumBohr:2010ta,Bern:2010ue,BjerrumBohr:2010hn,Vaman:2010ez,Bern:2010yg,Bern:1998sv,Bern:2006kd,ArkaniHamed:2008gz,BjerrumBohr:2006gg}. 
Part of this program is the extension of the
tree-level KLT theories, but many relations have been found at loop level as well. Since this work is extensive, we will not attempt to review it 
all here.
Since non-planar graphs appear on an equal footing with planar graphs in ${\cal N}=8$ 
supergravity, it seems important to understand non-planar graphs in ${\cal N}=4$ SYM if a weak-weak duality is to be explored. This is the focus 
of this section. 

We will review the known exact relations between the 4-point functions of subleading ${\cal N}=4$ SYM and those of ${\cal N}=8$ supergravity, at 
one and two-loops. For more than two-loops, the known relation for $n=4$ is for the leading IR singularity only. One application of these 
ideas for $n=5$ at one-loop is a new form of (tree level) KLT relations. 
Another are possible relations between ${\cal N}=4$ subleading-color amplitudes and ${\cal N}=8$ sugra for $n\geq 5$.

\subsection{One and two-loop relations}

In this subsection, we demonstrate the existence 
of some exact relations between 
$\cN=4$ SYM amplitudes  and
$\cN=8$ supergravity amplitudes at the one- and two-loop levels. 
The $L$-loop $N$-independent SYM amplitude $A^\EllL$ may be expected to be related to
the $L$-loop supergravity amplitude, 
as both have $\cO(1/\ep^L)$ leading IR divergences. 
Other subleading-color SYM amplitudes $A^\Ellk$ 
have $\cO(1/\ep^{2L-k})$ leading IR divergences, 
and consequently satisfy relations involving 
lower-loop supergravity amplitudes.\footnote{The normalization
of $\Asy^\Ellodd (s,t)$ is arbitrary.  
We have chosen one that is most natural in the context of
the SYM/supergravity relations presented in this subsection.}

In this section we use the notation
\be
\Asy^\Elleven (s,t) = a^L A^\Elleven_{[1]}, \qquad
\Asy^\Ellodd (s,t) = - {a^L \over \sqrt 2} A^\Ellodd_{[8]}
\ee
noting that the other components $A^\Ellk_{[i]}$ 
are obtained by permutations of  $s$, $t$, and $u$.
However, we omit the argument $(s,t)$ for functions that
are completely symmetric under permutations of $s$, $t$, and $u$.

Factor out the tree amplitude to define
\be
\Msy^\Ellk (s,t)=  {\Asy^\Ellk (s,t)\over \Asy^\Zero(s,t)} \, ,
\ee
so that the coupling constant $a^L$ is now included in the definition
of $\Msy^\Ellk (s,t)$, and where\footnote{In what follows we denote $A_{SYM}^{tree}(ij...k)=A(ij...k)$.} (see also (\ref{tree}))
\be
\Asy^\Zero(s,t)=-\frac{4iK}{st}
\ee

Recall that the one-loop $N$-independent SYM four-gluon amplitude is given by (\ref{Monedef}), obtaining
\be
\Asy^\OneDT =
- 2 \sqrt2 i K 
\left[\frac{g^2 N}{8\pi^2}\left(4\pi \e^{-\gamma}\right)^\ep\right] 
\left[ I_4^\One (s,t)+ I_4^\One (t,u)+ I_4^\One (u,s) \right]\,.
\ee
The one-loop supergravity four-graviton amplitude\footnote{after 
stripping off a factor of $(\kappa/2)^2$ for a four-point amplitude}
may be expressed as \cite{Green:1982sw,Bern:1998ug}
\be
\Asgone = 8iK^2 
\left[\frac{(\kappa/2)^2}{8\pi^2}\left(4\pi \e^{-\gamma}\right)^\ep\right] 
 \left[ I_4^\One (s,t)+ I_4^\One (t,u)+ I_4^\One (u,s) \right]\,.
\ee
The supergravity amplitude is proportional to $K^2$ rather than $K$ 
due to the KLT relations  \cite{Kawai:1985xq}
(a manifestation of the relation ``closed string = (open string)$^2$").
This factor is also present in the tree-level supergravity amplitude,
so we can factor it out
\be
\Asgone =    \Asgzero \Msgone   =\left(16iK^2\over stu\right) \Msgone
\ee
Defining $\lSYM = g^2 N$ and $\lSG = (\kappa/2)^2$,
one observes that the one-loop SYM and supergravity amplitudes are related by
\be
\Msy^\OneDT(s,t)= \sqrt2 \, \frac{\lSYM}{\lSG  u}\Msgone\,.
\label{oneloopMrelation}
\ee
In other words, 
the ratio of the one-loop subleading-color SYM 
and the one-loop supergravity amplitudes 
(after factoring out the tree amplitudes) 
is simply proportional to the ratio of coupling constants, 
where we encounter the effective dimensionless coupling $\lSG u$
for supergravity because $\lSG$ is dimensionful. 

Finally, rewrite \eqn{oneloopMrelation} in the manifestly
permutation-symmetric form
\be
\frac{1}{3}\left[(\lSG u)\Msy^\OneDT(s,t) + \rmcp \right]
=\sqrt2\lSYM\Msgone
\label{oneloopsugra}
\ee
(where $\rmcp$ denotes cyclic permutations of $s$, $t$, and $u$)
even though $u\Msy^\OneDT(s,t)$ is already symmetric under permutations.
A similar symmetrized relation can be written for the 
one-loop leading-color amplitude 
\be
(\lSG u)\Msy^\OneLC(s,t)+\rmcp 
=-\lSYM\Msgone
\label{onelooprelat}
\ee
obtained from the one-loop decoupling relation (\ref{oneloopdecouple}) 
together with \eqn{oneloopMrelation}. 

Now turn to {\it two loops}.
There are some relations between SYM and 
supergravity amplitudes that hold only for the IR-divergent terms.
The easiest case to analyze is the
two-loop $N$-independent SYM amplitude $\Asy^\TwoSC (s,t)$, 
since, from (\ref{twoloopmat})
\be
\Asy^\TwoSC (s,t)  =
- \sqrt2 a  \frac{X-Y}{\ep} \Asy^{\OneDT}(2\ep) + {\cal O}(\ep^0)\,.
\label{twosc}
\ee
This can be rewritten as 
\be 
\Msy^\TwoSC (s,t)
=
-2 a \,\frac{\lSYM}{\lSG u}
\left(\frac{X-Y}{\ep} \right)
\Msgone(2\ep)+{\cal O}(\epsilon^0)\, ,
\label{IRSC}
\ee
where $X=\log (t/u), Y=\log (u/s), Z=\log (s/t)$, as in (\ref{defXYZ}), 
thus obtaining a relation to the one-loop supergravity amplitude.

Using the relation 
$\Msgtwo (\ep) =\half [\Msgone(\ep) ]^2+ \cO(\ep^0)$ between the one- and two-loop supergravity 
amplitudes \cite{Naculich:2008ys,Brandhuber:2008tf,Weinberg:1965nx,Naculich:2011ry},
we can write this as  
\be
\frac{1}{3}
\left[(\lSG u)^2\Msy^\TwoSC(s,t) + \rmcp \right]
= 2\lSYM^2 \Msgtwo
\label{twoloopsugra}
\ee
where  this  relation is exact (!),
as may be easily verified by using the
exact expression  for the $N$-independent 
SYM amplitude \cite{Bern:1997nh} and from (\ref{twoloopSC})
\bea
\Msy^\TwoSC (s,t) 
&=& 
\frac{a^2 s t}{2} 
\Bigl[
s \left(I_4^\TwoP(s, t) + I_4^\TwoNP(s, t)
+ I_4^\TwoP(s, u)  + I_4^\TwoNP(s, u)\right)
\label{twoloopSG}\\ && \hspace{11mm}
+t \left(I_4^\TwoP(t, s) + I_4^\TwoNP(t, s)
+ I_4^\TwoP(t, u)  + I_4^\TwoNP(t, u)\right)
\nonumber\\ && \hspace{10mm}
-2u \left(I_4^\TwoP(u, s) + I_4^\TwoNP(u, s)
+ I_4^\TwoP(u,t)  + I_4^\TwoNP(u,t)\right)
\Bigr]  
\nonumber
\eea
and that for the two-loop supergravity amplitude \cite{Bern:1998ug}
\be
\Msgtwo 
= - \frac{s^3 t u }{4} 
\left[\frac{(\kappa/2)^2}{8\pi^2}
\left(4\pi \e^{-\gamma} \right)^\ep\right]^2
[I_4^{(2)P}(s,t)+I_4^{(2)NP}(s,t)
+I_4^{(2)P}(s,u)+I_4^{(2)NP}(s,u)]+ \rmcp
\ee
where $I_4^{(2)P}$ and $I_4^{(2)NP}$ are the two-loop planar and non-planar 4-point functions.

Now consider the two-loop subleading-color amplitude $\Msy^\TwoDT$. 
The two-loop decoupling relation 
(\ref{twoloopdecouple}) can be rewritten as 
\be
-\sqrt2 \left[ u\Msy^\TwoDT (s,t)+ \rmcp \right]
=6 \left[ u\Msy^\TwoLC (s,t) + \rmcp \right]\,.
\label{rela}
\ee
Using the ABDK relation \cite{Anastasiou:2003kj}
\be
\Msy^\TwoLC (\ep) =
\frac{1}{2}\left[\Msy^\OneLC (\ep) \right]^2
+ a f^\Two(\ep)\Msy^\OneLC (2\ep)+\cO(\ep), \qquad
f^\Two(\ep) = - (\zeta_2  + \ep \zeta_3 + \ep^2 \zeta_4)
\ee
together with \eqn{onelooprelat},
we can rewrite \eqn{rela} as
\bea
\frac{1}{3}\left[(\lSG u)\Msy^\TwoDT(s,t)+\rmcp \right]
&+&  \frac{1}{\sqrt2} \left\{
(\lSG u)\left[\Msy^\OneLC(s,t)\right]^2+ \rmcp \right\}
\nonumber\\
&=&
 \sqrt2 \frac{\lSYM^2}{8\pi^2} \left(4\pi \e^{-\gamma}\right)^\ep
 f^{(2)}(\ep)\Msgone(2\ep) +\cO(\ep)\,.
\label{twoloopre}
\eea
Unlike (\ref{twoloopsugra}), 
however, \eqn{twoloopre} only holds through $\cO(\ep^0)$,
which relates to the one-loop supergravity amplitude rather
than the two-loop one. 

Note that  (\ref{oneloopsugra}) and (\ref{twoloopsugra}) 
can be written as
\be
\frac{1}{3}
\left[(\lSG u)^L\Msy^\EllL(s,t) + \rmcp \right]
 \quad = \quad (\sqrt2\lSYM )^L \MsgEll 
\label{llloop}
\ee
for $L=0$, 1, and 2. Can this relation be valid at higher loops?
It turns out not to be the case, but we can still find some relations valid for $L\geq 3$.

\subsection{Three or more loops}

On the supergravity side,  there is an exact 
exponentiation formula \cite{Weinberg:1965nx,Naculich:2011ry},
which implies
\be
\MsgEll 
= \frac{1}{L!} \left[\Msgone  \right]^L 
+{\cal O}\left(\frac{1}{\epsilon^{L-2}}\right)
= \frac{1}{L!} \left[- \lSG (s Y - tX) \over  8 \pi^2 \ep \right]^L 
+{\cal O}\left(\frac{1}{\epsilon^{L-1}}\right)\,.
\label{sugrairdiv}
\ee
Since the leading IR divergences of $A^\EllL$ is $\cO(1/\ep^L)$, 
one can show that
the following relations hold:
\bea
&&\left[\lSG^2\frac{s^2+t^2+u^2}{3}\right]^k
\frac{1}{3}\left[(\lSG u) \Msy^{(2k+1,2k+1)} (s,t;\ep)+\rmcp\right]
\label{2k1}\\
&&\hspace{-1.6cm}
=\lSYM^{2k+1} \frac{2^{2k+1/2}}
{(2k+1)!}
\left[\Msgtwo (\ep) +\frac{1}{6} 
\left( \lSG \over 8 \pi^2 \right)^2 
\left(\frac{sX+tY+uZ}{\ep}\right)^2
\right]^k
\Msgone (\ep)
+{\cal O}\left(\frac{1}{\epsilon^{2k}}\right)
\nonumber
\eea
for $L=2k+1$ and
\bea
&&
\left[\lSG^2 \frac{s^2+t^2+u^2}{3}\right]^k
\frac{1}{3}
\left[(\lSG u)^2 \Msy^{(2k+2,2k+2)} (s,t;\ep)+\rmcp \right]
\label{2k2} \\
&&\hspace{-1.9cm}
=\lSYM^{2k+2} \frac{2^{2k+2}}{(2k+2)!}
\left[\Msgtwo (\ep)+\frac{1}{6}
\left( \lSG \over 8 \pi^2 \right)^2 
\left(\frac{sX+tY+uZ}{\ep}\right)^2
\right]^k \Msgtwo (\ep)
+{\cal O}\left(\frac{1}{\epsilon^{2k+1}}\right)
\nonumber
\eea
for $L=2k+2$ (where $k=0,1,2,...$).

That is, we have an exact relation at $L-$loops for the leading IR
divergence $\sim {\cal O}(1/\epsilon^L)$, with an untested relation
for the subleading divergence of ${\cal O}(1/\epsilon^{L-1})$.
See also \eqn{mostsubleading}.

An interesting fact is that either \eqn{llloop} 
(or \eqns{2k1}{2k2} without the extra term), 
and also the relation (\ref{twoloopre}),  
have a possible interpretation in terms of the 
't Hooft string picture of the $1/N$ expansion. 
Thus at least in the case of $L=1,2$,  \eqns{llloop}{twoloopre} 
still do, so one can hope that there is a
correct relation at higher $L$ yet to be determined.

\subsection{New KLT relations}

One of the pioneering connections between SYM and supergravity theories 
are the KLT relations \cite{Kawai:1985xq}, originally proved using string theory methods \cite{Kawai:1985xq,Berends:1988zp}. 
More recently, alternate versions of KLT relations have been presented based on field theoretic techniques at the tree level
\cite{BjerrumBohr:2010ta,BjerrumBohr:2010yc}. One form of 
these new relations has manifest $(n-3)!$ permutation symmetry for the $n$-point functions, and another has $(n-2)!$ symmetry, but requires 
regularization as a consequence of singularities.  They are part of a flurry of recent activity
relating ${\cal N}=4$ SYM and ${\cal N}=8$ supergravity, including 
\cite{Elvang:2007sg,Ananth:2007zy,Feng:2010br,Bern:2010ue,BjerrumBohr:2010hn,Bern:2008qj,Vaman:2010ez,Bern:2010yg}
(among older works see also \cite{Bern:1998sv,Bern:1994zx,Bern:1996ja}).
Recent work applying the KLT relations include \cite{BjerrumBohr:2010zb,Feng:2010hd,Tye:2010kg,Elvang:2010kc}.
In our quest for SYM-supergravity relations, we first review previous KLT relations; we then note
 that $A_{5;3}$ and the 1-loop 
supergravity amplitude both have $1/\epsilon$ IR divergences.
We present here a tree-level KLT relation for the $n=5$-point amplitudes derived in \cite{Nastase:2010xa}, 
using information from one-loop SYM and supergravity 
amplitudes and their IR divergences. This results in a KLT relation for 5-point functions with $2(n-2)!$ manifest symmetry, without the need 
for regularization. These KLT relations are proved explicitly using the helicity spinor formalism and the Parke-Taylor formula. 
In analogy with section 3.1
on 4-point functions of ${\cal N}=8$ supergravity and subleading-color ${\cal N}=4$ SYM theories, 
both with the $1/\epsilon$ IR divergence, we explore the possibility that the 1-loop 5-point supergravity amplitude can be expressed as a linear 
combination of the $A_{5;3}$ SYM amplitudes. In particular a linear relation is proposed among the $1/\epsilon$ IR divergences of the two theories. 

At tree  level, the KLT relations are quadratic relations between the $n$-point amplitudes of ${\cal N}=4$ SYM and those of 
${\cal N}=8$ supergravity. In these relations, the helicity information is all contained within the amplitudes, and the 
coefficients are all function of the kinematic invariants $s_{ij}$ only.

These relations relate graviton tree amplitudes with sums of squares (products) of gauge tree amplitudes. 
The original KLT relations were derived from string theory in the $\a'\rightarrow 0$ limit \cite{Kawai:1985xq,Berends:1988zp}, and 
can be expressed as (we use the notation of \cite{Bern:1998sv})
\bea
A^{tree}_{n,sugra}(12...n)&=&(-1)^{n+1}\Big[A_n(12...n)\sum_{{\rm perms}}f(i_1...i_j)\bar f(l_1...l_{j'})\times\cr
&&\times A_n(i_1,...,i_j,1,n-1,l_1,...,l_{j'},n)\Big]+{\cal P}(2,...,n-2)\cr
f(i_1,...,i_j)&=&s(1,i_j)\prod_{m=1}^{j-1}\Big(s(1,i_m)+\sum_{k=m+1}^jg(i_m,i_k)\Big)\cr
\bar f(l_1,...,l_{j'})&=&s(l_1,n-1)\prod_{m=2}^{j'}\Big(s(l_m,n-1)+\sum_{k=1}^{m-1}g(l_k,l_m)\Big)
\eea
where ``perms'' are $(i_1,...,i_j)\in {\cal P}(2,...,n/2)$, $(l_1,...,l_{j'}\in{\cal P}(n/2+1,...,n-2)$, $j=n/2-1,j'=n/2-2$ and 
$g_{i,j}=s_{ij}$ if $i>j$ and zero otherwise.

In \cite{BjerrumBohr:2010ta} and \cite{BjerrumBohr:2010yc}, new forms of the KLT relations for any $n$-point function were found. 
They are both written in terms of the functions 
\bea
&&{\cal S}[i_1...i_k|j_1...j_k]=\prod_{t=1}^k(s_{i_t1}+\sum_{q>t}^k\theta(i_t,i_q)s_{i_ti_q})\cr
&&\tilde {\cal S}[i_1...i_k|j_1...j_k]=\prod_{t=1}^k(s_{j_tn}+\sum_{q<t}^k\theta(j_q,j_t)s_{j_qj_t})
\eea
where $\theta(i_t,i_q)$ is zero in $(i_t,i_q)$ has the same order in both sets ${\cal I}=\{i_1,...,i_k\}$ and 
${\cal J}=\{j_1,...,j_k\}$ and is 1 otherwise, and similarly for $\theta(j_q,j_t)$. 

A form of KLT relations was found in \cite{BjerrumBohr:2010ta}, but need to be regularized,
due to a singular denominator,
\bea
A^{tree}_{n,sugra}&=&(-1)^n \sum_{\gamma, \b}\frac{ {\tilde A}_n(n,\gamma_{2,n-1},1){\cal S}[\gamma_{2,n-1},\beta_{2,n-1}]_{p_1}
A_n(1,\beta_{2,n-1},n)}{s_{12...n-1}}\cr
A^{tree}_{n,sugra}&=&(-1)^n\sum_{\b, \gamma}\frac{  A_n(n,\b_{2,n-1},1)\tilde {\cal S}[\b_{2,n-1},\gamma_{2,n-1}]_{p_n}
\tilde A_n(1,\gamma_{2,n-1},n)}{s_{23...n}}\label{KLTold}
\eea
however they have a large $(n-2)!$ manifest symmetry. Another set was proven in \cite{BjerrumBohr:2010yc} which is non-singular,
\bea
A^{tree}_{n,sugra}&=&(-1)^{n+1}\sum_{\sigma\in S_{n-3}}\sum_{\a\in S_{j-1}}\sum_{\b\in S_{n-j-2}}A_n(1,\sigma_{2,j},
\sigma_{j+1, n-2}, n-1,n){\cal S}[\a_{\sigma(2),\sigma(j)}|\sigma_{2,j}]_{p_1}\cr
&&\times \tilde {\cal S}[\sigma_{j+1,n-2}|\b_{\sigma(j+1),\sigma(n-2)},n]_{p_n}\tilde A_n(\a_{\sigma(2),\sigma(j)},
1,n-1,\b_{\sigma(j+1),\sigma(n-1)},n)\label{KLTbohr}
\eea
but with only $(n-3)!$ manifest symmetry.

The original KLT relation for the 5-point function is  
\be
A^{tree}_{5,sugra}=s_{12}s_{34}A(12345)\tilde A(21435)
+s_{13}s_{24}A(13245)\tilde A(31425)\label{usualKLT}
\ee
and has $(n-3)!=2!$ symmetry, whereas the KLT relations (\ref{KLTbohr}) become, explicitly,
\bea
A^{tree}_{5,sugra}&=&\sum_{\sigma,\tilde \sigma\in S_2}\tilde A(45,\tilde\sigma_{23},1)A(1,\sigma_{23},45)S[\tilde \sigma_{2,3}|\sigma_{2,3}]_{p_1}\cr
&=&s_{12}s_{13}(A(45231)A(12345)+A(45321)A(13245))+s_{13}(s_{12}+s_{23})A(45231)A(13245)\cr
&&+s_{12}(s_{13}+s_{23})A(45321)A(12345)\cr
A^{tree}_{5,sugra}&=&\sum_{\sigma,\tilde \sigma\in S_2}\tilde A(14,\tilde\sigma_{23},5)A(1,\sigma_{23},45)\tilde S[\sigma_{2,3}|\tilde 
\sigma_{2,3}]_{p_4}\cr
&=&s_{24}s_{34}[A(12345)A(14235)+A(13245)A(14325)]+
s_{34}(s_{24}+s_{23})A(12345)A(14325)\cr
&&+s_{24}(s_{34}+s_{23})A(13245)A(14235)\label{KLT5}
\eea 
and have $(n-3)!=2!$ symmetry. 

We now derive another KLT relation for 5-point amplitudes using information about subleading one-loop amplitudes. 

As we saw in 
(\ref{subleadingrel}), the $A_{n;j}$ are related to the $A_{n;1}$ via group theory.
In particular, for 5-point amplitudes, one has a single-trace amplitude $A_{5;1}$ and a double-trace
amplitude $A_{5;3}$ related by \cite{Bern:1991aq}
\be
A_{5;3}(45123)=\sum_{\sigma\in COP_4^{123}}A_{5;1}(\sigma(1),...,\sigma(4),5)\label{12trace}
\ee

The single-trace amplitude is given by
\be
A_5^{(1,0)}(12345)\equiv A_{5;1}(12345)=-\frac{1}{4}A(12345)\sum_{\cyclic}F^{(1)}(s,t,m^2)
\ee
where
\be
F^{(1)}(s,t,m^2)=st I^{(1)}_5 (s,t,m^2)
\ee
is the dimensionless one-mass box, and $I^{(1)}(s,t,m^2)$ is the 1-loop scalar box integral (\ref{Monedef}) with momenta 3,4 in the same corner
and $m^2=P^2=(p_3+p_4)^2$.

Substituting in (\ref{12trace}), we find
\be
A_{5;3}(fg;hij)=\sum_{abcde\in 30 \; {\rm fixed \; terms}}F(cde;ab)[s_{abcde;+;fghij}A(abcde)+s_{abcde;-;fghij}A(abedc)]
\ee
Here $s_{abcde;\pm;fghij}$ are signs, defined as follows. The relative sign is plus if $ab$ belong to $hij$, 
and minus otherwise, and the overall sign is plus if the permutation of $hij$ inside $abcde$ is even, and minus 
if it is odd.

The 1-loop ${\cal N}=8$ supergravity amplitude is \cite{Bern:2006kd}, written in terms of the scalar 1m box 
$I(123,45)$ (with momenta 4,5 on the same corner of the box) and 
the dimensionless box $F(123;45)$ is 
\be
A_{5,sugra}^{1-loop}(12q_3q_2q_1)
=-\frac{1}{2}\sum_{30\; {\rm perms}}s_{q_2q_1}s_{12}s_{2q_3}A(12q_3q_2q_1)A(12q_3q_1q_2)F(12q_3;q_2q_1)\label{intermediat}
\ee
or
\be
A_{5,sugra}^{1-loop}(12345)=-\frac{1}{2}\sum_{30\; {\rm perms}}F(cde;ab)s_{cd}s_{de}s_{ab}A(cdeab)A(cdeba)\label{1loopsugra}
\ee

The IR behavior of the 1-loop 1m scalar box is
\bea
I_{4,1m}(s,t,m^2)&=&\frac{r_\Gamma}{s_{12}s_{23}}\Big\{\frac{2}{\epsilon^2}[(-s_{12})^{-\epsilon}
+(-s_{23})^{-\epsilon}-(-s_{45})^{-\epsilon}]+{\rm finite}\Big\}\Rightarrow\cr
F(cde;ab)&\simeq& \frac{r_\Gamma}{\epsilon^2}[(-s_{cd})^{-\epsilon}+(-s_{de})^{-\epsilon}-(-s_{ab})^{-\epsilon}]
+{\rm finite}\cr
r_\Gamma&=&\frac{\Gamma(1+\epsilon)\Gamma^2(1-\epsilon)}{\Gamma(1-2\epsilon)},\label{IRscalarbox}
\eea
where $D=4-2\epsilon$.

{\bf IR behavior of the double-trace 1-loop SYM amplitude $A_{5;3}$}. 

Using (\ref{IRscalarbox}), we find
\bea
A_{5;3}(fg;hij)&=&\sum_{abcd\in 30\;{\rm terms}}F(cde;ab)[s_{abcde;+;fghij}A(abcde)+s_{abcde;-;fghij}A(abedc)]\cr
&\sim&\frac{r_\Gamma}{\epsilon^2}\sum_{abcd\in 30\;{\rm terms}}
[s_{cd}^{-\epsilon}+s_{de}^{-\epsilon}-s_{ab}^{-\epsilon}][s_{abcde;+;fghij}A(abcde)\cr
&&+s_{abcde;-;fghij}A(abedc)]
\eea
Organizing the coefficients of each divergence we find
\be
A_{5;3}(fg;lmn)\simeq \frac{r_\Gamma}{\epsilon^2}\sum_{i<j}(-s_{ij})^{-\epsilon}\sum_{abc\neq i,j}\epsilon_{lmn}[A(ijabc)]
\ee
where $\epsilon_{lmn}[A(ijabc)] $ means $A(ijabc)$ is multiplied by the sign of the permutation of $l,m,n$ inside 
$i,j,a,b,c$, and the sum over $a,b,c$ contains all the 6 terms of the arbitrary permutation of the $a,b,c\neq i,j$.

The leading ($1/\epsilon^2$) divergence of $A_{5;3}(45;123)$, given by
\be
\sum_{i<j}\sum_{abc\neq i,j}\epsilon_{123}[A(ijabc)]\label{leadingIRgen}
\ee
vanishes by explicit computation, so that
the leading IR divergence of $A_{5;3}$ is $1/\epsilon$, as expected from a generalization of the subleading-color amplitude of the 4-gluon 
amplitude \cite{Naculich:2008ew,Naculich:2009cv}.\footnote{The vanishing of the $1/\epsilon^2$ IR divergence of (\ref{leadingIRgen}) is also a 
consequence of (\ref{5pointKK}).}

{\bf IR behavior of ${\cal N}=8$ supergravity one-loop amplitudes and KLT relations}. 

Using (\ref{intermediat}), we obtain 
\bea
A_{5,sugra}^{1-loop}(12345)&=&-\frac{1}{2}\sum_{30\; {\rm perms}}F(cde;ab)s_{cd}s_{de}s_{ab}A(cdeab)A(cdeba)\cr
&\simeq & -\frac{1}{2\epsilon^2}\sum_{30\; {\rm perms}}[s_{cd}^{-\epsilon}+s_{de}^{-\epsilon}-s_{ab}^{-\epsilon}]s_{cd}s_{de}s_{ab}A(cdeab)A(cdeba)
\label{5sugraIRbeh}
\eea
Organizing the terms by IR divergences, we obtain
\bea
&&A_{5,sugra}^{1-loop}\simeq \frac{1}{\epsilon^2}\sum_{i<j}s_{ij}^{1-\epsilon}
\times \Big[\sum_d s_{cd}s_{de}A(ijcde)A(ijedc)\cr
&&+\sum_c s_{ic}s_{ab}A(ijabc)A(ijbac)+\sum_c 
s_{jc}s_{ab}A(ijcba)A(ijcab)\Big]
\eea

On the other hand, we know that the IR behavior of the 1-loop $n-$point supergravity amplitude is 
\cite{Dunbar:1995ed}
\be
A_{n,sugra}^{1-loop}(1...n)\simeq \frac{1}{\epsilon^2}A_{n,sugra}^{tree}(1...n)\sum_{i<j} s_{ij}^{1-\epsilon}
\ee
which means that we must have the KLT relation
\bea
A_{5,sugra}^{tree}(12345)&=&\sum_d s_{cd}s_{de}A(ijcde)A(ijedc)+\sum_c s_{ic}s_{ab}A(ijabc)A(ijbac)\cr
&&+\sum_c s_{jc}s_{ab}A(ijcba)A(ijcab),\hspace{3cm}
\forall (ij)\label{KLTus}
\eea
Note that it has the larger manifest symmetry of $2\times (n-2)!=2\times 3!$, and has no need to be regularized.

The tree-level KLT formula (\ref{KLTus}) has been derived using the IR behavior of 1-loop computations. However it can be proved explicitly. 
To do so,  use the helicity spinor formalism and the Parke-Taylor formula \cite{Parke:1986gb}, which states that 
\be
A_{n,SYM}^{tree}(1^+2^+...i^-...j^-...n^+)=\frac{\langle ij\rangle ^4}{\langle 12\rangle\langle23\rangle...\langle n1\rangle}
\ee
or for our case, for instance choosing $1^-2^-$,
\be
A(1^-2^-3^+4^+5^+)=\frac{\langle12\rangle^4}{\langle12\rangle\langle23\rangle\langle34\rangle\langle45\rangle\langle51\rangle}
\ee
A similar formula exists for the supergravity amplitude \cite{Bern:1998sv}
\be
A_{5,sugra}^{tree}(1^-2^-3^+4^+5^+)=\frac{\langle 12\rangle^8\epsilon(1234)}{N(5)}\label{5sugra}
\ee
where 
\bea
\epsilon(ijkl)&=&4i\epsilon_{\mu\nu\rho\sigma}k_i^\mu k_j^\nu k_k^\rho k_l^\sigma\cr
N(5)&=&\prod_{i=1}^4\prod_{j=i+1}^5\langle ij\rangle
\eea

A specific case of (\ref{KLTus}) is proved, namely
\bea
&&A_{5,sugra}^{tree}(12345)\cr
&=&s_{34}s_{45}A(12345)A(12543)+s_{53}s_{34}A(12534)A(12435)+s_{45}s_{53}A(12453)A(12354)\cr
&&+s_{23}s_{45}A(12345)A(12354)+s_{24}s_{35}A(12435)A(12453)+s_{25}s_{34}A(12534)A(12543)\cr
&&+s_{13}s_{45}A(21345)A(21354)+s_{14}s_{35}A(21435)A(21453)+s_{15}s_{34}A(21534)A(21543)\cr
&&\label{KLTnew}
\eea
The others follow from permutations and symmetry.

One makes use of helicity spinor identities
to verify that the right hand side of (\ref{KLTnew}) is equal to (\ref{5sugra}), proving the KLT relation. 

{\bf Relation between $A_{5,sugra}^{one-loop}$ and $A_{5;3}$}.

Motivated by the fact that the leading IR divergence of the $n=5$-point  supergravity amplitude and of $A_{5;3}$ are both of order $1/\epsilon$
at 1-loop,  one investigates whether $A_{5,sugra}^{1-loop}$ can be expressed as a linear combination of $A_{5;3}$ amplitudes. One 
uses information from eqs. (\ref{IRscalarbox}) to (\ref{5sugraIRbeh}), 
and finds a relation valid for IR divergences, and then one conjectures how one possibly could extend to a relation for the full amplitudes.

Based on what happened at $4-$points at $1-$ and $2-$loops, as discussed in section 3.1, we want to find 
$A_{5,sugra}^{1-loop}$ as a linear combination of the $A_{5;3}$ amplitudes.

In analogy with with the 4-point function, we would like to find a relation of the type
\bea
&&A^{1-loop}_{5,sugra}(12345)=\sum_{i\in fghij} \beta_i A_{5;3}(i)\cr
&&=\sum_{abcde\in 30\; {\rm fixed\; terms}}F(cde;ab)
\sum_{i\in fghij}\beta_i[s_{abcde;+;i}A(abcde)+s_{abcde;-;i}
A(abedc)]\cr
&&\label{type}
\eea
On the other hand, 
\be
A^{1-loop}_{5,sugra}(12345)=\sum_{abcde\in 30\; {\rm fixed \; terms}}F(cde;ab)\a_{abcde}
\ee
where from (\ref{1loopsugra}), 
\be
\a_{abcde}=-\frac{1}{2}s_{cd}s_{de}s_{ab}A(cdeab)A(cdeba)
\ee
which means that we need
\be
\a_{abcde}=\sum_{i\in fghij}\beta_i[s_{abcde;+;i}A(abcde)+s_{abcde;-;i}
A(abedc)]\label{betaeq}
\ee
to be satisfied,
which are 30 equations for 10 unknowns ($\beta_i$), so (\ref{betaeq}) are not guaranteed to have solutions.

The 30 equations can then be rewritten, using the explicit form of $\a_{abcde}$, and a new notation that will 
prove useful, as
\be
-\frac{1}{2}s_{ab}s_{bc}s_{de}A(abcde)A(abced)=\sum_{fg;hij}\beta_{(fg)}\epsilon_{hij}[A(abcde)]\Big(1-\epsilon_{hij}(de)\frac{A(abced)}{A(abcde)}\Big)
\label{betaeqs}
\ee
where $\epsilon_{hij}(de)$ is plus if both $d,e$ belong to $h,i,j$, and minus otherwise.

In order to see if a unique solution for the $\b_i$ is possible, one can match the IR behaviors on the two sides of 
(\ref{type}). Expressing the IR behaviors of the lhs and the rhs, 
\be
\frac{1}{\epsilon^2}A_{5,sugra}^{tree}(12345)\sum_{i<j}s_{ij}(-s_{ij})^{-\epsilon}=\frac{r_\Gamma}{\epsilon^2}\sum_{k\in fg;lmn}\beta_k\sum_{i<j}
(-s_{ij})^{-\epsilon}\sum_{abc\neq i,j}\epsilon_{lmn}[A(ijabc)]
\ee
which means that one requires, using the vanishing of the $1/\epsilon^2$ IR divergence,
\be
A_{5,sugra}^{tree}(12345)s_{ij}=\sum_{k\in fg;lmn}\beta_k\sum_{abc\neq i,j}\epsilon_{lmn}[A(ijabc)]
\ee
If we denote the $A_{5,sugra}^{tree}(12345)$ by just $M_5$, then the lhs is a vector column of $(ij)$, $M_5s_{ij}$. Also denote $\sum_{abc\neq i,j}\epsilon_{lmn}[A(ijabc)]$ as $N_{(ij),(fg)}$, so that
\be
N_{(ij),(fg)}\beta_{(fg)}=M_5 s_{ij}\Rightarrow [\beta_{(fg)}]=[N_{(ij),(fg)}]^{-1}M_5s_{ij}\label{coeff}
\ee
Note that the index $(fg)$ on the matrix $N$ has 10 values, and these values can also be identified by the $lmn$ of $\epsilon_{lmn}[A(ijabc)]$, since
it corresponds to the same 10 terms, picking out a group $(fg)$ or $(lmn)$ out of $1,2,3,4,5$.

At this point however note that the vanishing of the leading IR divergence in (\ref{leadingIRgen}) means that 
\be
\sum_{(ij)}N_{(ij),(fg)}=0
\ee
i.e., that the matrix $N$ has rank 9 instead of 10.  One then needs to work with the corresponding 
$9\times 9$ reduced matrix $N_{red;(ij),(fg)}$ and give the 10th coefficient $\beta_{(fg)}$ an arbitrary value.

Therefore one has found a linear relation, (\ref{type}), with coefficients obtained from (\ref{coeff}), which is satisfied by the IR divergences, 
and containing an arbitrary parameter. 
Of course, it is still not clear that the remaining $\b_{(fg)}$ are unique. For that, one must calculate the rank of $N_{red}$. 
If its rank is less than 9, the solution is parametrized by more than 
one parameter, since then some of the remaining $\beta$'s will be undetermined. As the algebra is quite involved, 
this is a project for further work.

In order to see if (\ref{type}) is also satisfied for the full amplitude, 
one must substitute the solution for $\b_{(fg)}$ back in (\ref{betaeqs}) and see if these equations are satisfied, since now one needs to check 
whether the 30 equations are satisfied by substituting the 10 unknowns $\b_{(fg)}$ solved as in (\ref{coeff}). 
The verification of (\ref{type}) for $n=5$ is analogous with that for the (successful) relation (\ref{onelooprelat}) for $n=4$. Therefore, it would 
be interesting if (\ref{type}), (\ref{betaeq}) were true.\footnote{As (\ref{llloop}) exemplifies for $L=1,n=4$, (\ref{type}) may not be the only 
equation relating ${\cal N}=8$ supergravity to ${\cal N}=4$ SYM for $L=1,n=5$.}

In principle the strategy described above can be applied to higher $n-$point amplitudes. 
Namely one can analyze the IR behavior of the results for ${\cal N}=4$ SYM and ${\cal N}=8$ supergravity 
at 1-loop, and compare these to the known behavior, which would imply a relation among tree amplitudes from SYM, and a 
KLT-type relation from the supergravity. Finally, one can relate the subleading-color SYM and supergravity 
amplitudes, and use the consistency of the IR behavior to fix the proposed relation. For $n=6, L=1$, the results of 
Bjerrum-Bohr, Dunbar, and Ita \cite{BjerrumBohr:2006gg} are suitable for this purpose.

\section{Geometric interpretations of subleading-color amplitudes}

\subsection{Polytope picture}

{\bf Polytopes for $MHV$ leading amplitudes}

In \cite{Mason:2010pg}, a simple picture was found for the 1-loop color ordered leading amplitudes of ${\cal N}=4$ SYM theory, 
in terms of the volume of a closed polytope in $AdS_5$. In \cite{Nastase:2011mx}, it was generalized to subleading-color amplitudes. 
The picture for the leading $MHV$ amplitude was obtained as follows. 
We start by writing the amplitude in a space dual to momenta, thus trivializing momentum conservation $\sum_i p_i=0$, by $p_i=x_i-x_{i+1}$.
Then, e.g., the 1-loop dimensionless massless box function (in 4 dimensions, 
which is of course IR divergent) 
$F_{0m}(1234)=-{1 \over 2} stI_4^{(1)}(s,t)$, with $I_4^{(1)}(s,t)$ in (\ref{Monedef}) becomes 
\be
F_{0m}(1,2,3,4)=i\int \frac{d^4x_0}{2\pi^2} \frac{(x_1-x_3)^2(x_2-x_4)^2}{(x_0-x_1)^2(x_0-x_2)^2(x_0-x_3)^2(x_0-x_4)^2}\label{box}
\ee

We then construct $x_{\a\dot\a}=x^\mu (\sigma_\mu)_{\a\dot\a}$ and finally map 
\be
x^{\a\dot\a}\rightarrow X^{AB}=\begin{pmatrix}-\frac{1}{2}\epsilon^{\a\b}x^2&ix^\a_{\dot\b}\\-ix^\b_{\dot\a}&\epsilon_{\dot\a\dot\b}\end{pmatrix}
\label{xtoX}
\ee
Here the $X$'s, satisfying
\bea
&&X^2\equiv \frac{1}{2}\epsilon_{ABCD}X^{AB}X^{CD}=0\cr
&& X_i\cdot X_j =-(x_i-x_j)^2,
\eea
are coordinate patches on the quadric $X\cdot X=0$ in $RP^5$, with $X^{AB}\sim \lambda X^{AB}$ their homogeneous coordinates.
These $X$'s are considered as vertices situated at the boundary of an $AdS_5$ and are simple bitwistors living in twistor space, i.e. 
$\exists$ twistors $A^A$ and $B^B$ such that $X^{AB}=A^{[A}B^{B]}$ (a twistor $A^A$ is made of $(A^\a,A_{\dot\a})$). 

Consider a box function characterized by vertices $X_1,X_2,X_3,X_4$. Then 
the following  function of the Feynman parameters $\a_i\in (0,1)$ with $\sum \a_i=1$,
\be
X(\a)=\a_1X_1+\a_2X_2+\a_3X_3+\a_4X_4
\ee
is a map to $RP^5$, but such that $X(\a)\cdot X(\a)\neq 0$, and in fact they vary over a tetrahedron in $RP^5$.
After normalizing by 
\be
Y(\a)=\frac{X(\a)}{\sqrt{X(\a)\cdot X(\a)}}
\ee
one obtains $Y(\a)\cdot Y(\a)=1$, which means that $Y(\a)$ lies in Euclidean $AdS_5$. Since straight lines $X(\a)$ are mapped to geodesics in $AdS_5$, 
the edges and faces of the tetrahedron in $AdS_5$ are geodesic, which by definition makes the tetrahedron {\em ideal}.

The value of the IR-finite 4-mass box  matches twice the volume of the tetrahedron in $AdS_5$. The IR divergent lower mass functions 
need to be regularized, either in dimensional regularization, or using a mass regularization as in \cite{Mason:2010pg}, 
modifying $X^2=0$ to $X\cdot X= \mu^2 (X\cdot I)$, with $I$ a fixed point (A useful choice of $I$ is $X_i\cdot I=1, \forall i$). 

The one loop MHV $n-$point amplitudes divided by the tree MHV amplitudes are given by the sum of 1-mass and 2-mass easy box functions with 
coefficient one, which add up to the volume of a closed 3-dimensional polytope (without a boundary) with $n$ vertices. 

Note that by this definition, the volume of a tetrahedron comes with a sign, determined by the order of the dual space vertices $x_i$ in  
the box function $F(i,j,k,l)$. That also induces an orientation (sign) for the triangular faces of the tetrahedron, determined by whether the 
missing vertex from $(ijkl)$ is in an even or odd position. Faces with same vertices and different orientation (sign) can be glued together, 
forming a continuous object.

{\bf Polytopes for $MHV$ subleading-color amplitudes}

For subleading-color amplitudes, we want to use (\ref{subleadingrel}) to relate to the leading amplitudes, and expand in the KK basis (\ref{KK}), where 
we will obtain a nice geometrical interpretation. 

We start with the 5-point amplitude as an example. The ratio of the leading 1-loop $MHV$ to the tree level $MHV$ amplitudes is the volume of a 
boundary of a 4-simplex, 
\be
\frac{A_{5;1}^{MHV}(12345)}{A_5^{MHV}(12345)}\equiv M_5^{MHV}(12345)=\sum_{\cyclic}I(x_1,x_2,x_3,x_4,(x_5))\equiv V(x_1,x_2,x_3,x_4,x_5)
\ee
Here $I(x_1,x_2,x_3,x_4,(x_5))$ is the volume of the tetrahedron with vertices $x_1,x_2,x_3,x_4$, equal to $F(1,2,3,4)$, and the missing vertex
$(x_5)$ is added 
in brackets since the cyclicity involves all 5 points; $V(x_1,x_2,x_3,x_4,x_5)$ is the volume of the boundary of the 4-simplex in twistor space,
with $(y)_i\rightarrow (Y)_i$, i.e. we map the arguments of $V$ into twistor space. 

Using (\ref{subleadingrel}) and writing the tree amplitudes in terms of the KK basis, we obtain 
\bea
A_{5;3}(12345)&=&A_5(12345)[(M_5(12345)-M_5(41235))+(M_5(43125)-M_5(31245))]\cr
&&+A_5(12435)[(M_5(12435)-M_5(31245))+(M_5(34125)-M_5(41235))]\cr
&&+A_5(14235)[(M_5(14235)-M_5(31425))+(M_5(34125)-M_5(41235))]\cr
&&+A_5(13245)[(M_5(23145)-M_5(42315))+(M_5(43125)-M_5(31245))]\cr
&&+A_5(13425)[(M_5(23145)-M_5(31425))+(M_5(43125)-M_5(24315))]\cr
&&+A_5(14325)[(M_5(23145)-M_5(31425))+(M_5(34125)-M_5(23415))]\cr
&& \label{5pointKK}
\eea
We see that for each KK basis member we have the sums of two terms which are some simple differences of $M_5$'s. In fact these differences can 
be written as the differences of two polytope volumes, which in turn can be written as the volume of a simple polytope. For instance, 
the coefficient of the KK basis element $A_5(12345)$ in (\ref{5pointKK}) is 
\bea
&&(M_5^{MHV}(12345)-M_5^{MHV}(41235))+(M_5^{MHV}(43125)-M_5^{MHV}(31245))\cr
&=&[V(x_1,x_2,x_3,x_4,x_5)-V((x_4-x_5+x_1),x_1,x_2,x_3,x_4)]\cr
&&+[V(x_4,(x_1+x_4-x_5),x_1,(x_1-x_3+x_4),(x_2-x_3+x_4))\cr
&&-V(x_1,(x_1-x_3+x_4),(x_2-x_3+x_4),x_4,x_5)]
\eea
In the two brackets, the two volumes of opposite sign correspond to polytopes with $n-1=4$ points common out of $n=5$, and the relative sign is 
such that we can write this as the volume of the polytope obtained by the taking the union of the two polytopes. 

By examining the $n=6$ case \cite{Nastase:2011mx} as well, we can understand the general pattern for $A_{n;3}$. The general formula is 
\bea
A_{n;3}^{MHV}(n-1,n,1,2,....,n-2)&=&\sum_{\{\sigma\}_i\in OP(\{\a\},\{\b^T\})|_{j_{max}}}A_n^{MHV}(1,\{\sigma\}_i,n)\cr
&&\times \sum_{n-1\in \{\a\},\{\b\};j_{max}\in\{\a\},\{\b\}}(-)^{n_\b}M_n^{MHV}(\{\b\},1,\{\a\},n)\cr
&&\label{an3}
\eea
with $M_n^{MHV}$ being the volume of a closed polytope, and pairs of opposite sign $M_n$'s adding up to another closed polytope.
It is obtained as follows. In the above, just from (\ref{subleadingrel}), we have
$M_n(\{\b\},1,\{\a\},n)$, where $n-1$ is either in $\{\a\}$ or in $\{\b\}$, and 
otherwise $\{\a\}$ contains $2,3,...,k$ and $\{\b\}$ contains $k+1,...,n-2$. 

For the tree amplitude prefactors, when using the KK relations (\ref{KK}), from 
$\{\a\}$ and $\{\b\}$ we form the permutation $\{\sigma\}_i$ which contains $\{\a\}$ and $\{\b^T\}$, keeping the ordering, i.e.
in the KK basis amplitude we have $A(1,\{\sigma\},n)$. Here if we extract the $n-1$, then $\{\a\}=2,...,j_{max}$ is ordered, i.e. it goes from 
left to right in the permutation, and then $\{\b\}=j_{max}+1,...,n-2$ is transposed and still ordered, i.e. it goes from right to left. 
The same $j_{max}$ (extracted from the resulting KK basis member) is obtained from either $k$ or $k+1$. That means that there are 
exactly 4 terms corresponding to the same KK basis member, corresponding to both $j_{max}$ and $n-1$ belonging to either $\{\a\}$ or $\{\b\}$.

The sign of the terms is obtained from the sign in the KK relations (\ref{KK}), i.e. $(-1)^{n_\b}$, where here $\{\b\}$ refers to the 
individual $M_n(\{\b\},1,\{\a\},n)$ term. The fact that $j_{max}$ belongs to either $\{\a\}$ or $\{\b\}$ means 
that in $M_n(\{\b\},1,\{\a\},n)$ we either have $j_{max}$ at the end of $\{\a\}$, or at the beginning of $\{\b\}$, i.e. we have 
a flip of $j_{max}$ $n$ vs. $n$ $j_{max}$ in between terms with different signs, hence a different $n_\b$ (with or without $j_{max}$). 
The exception is when actually $(n-1)$ is at the end of $\{\a\}$, and not $j_{max}$, 
in which case the same flip is now $(n-1)n$ vs. $n(n-1)$, and the same relative minus
sign applies. 

Since the pair in the difference  in the () bracket multiplying KK basis members
has the same $n-2$ permutation, and the remaining two terms are flipped, we have
the difference of two $n-$polytopes with a common $n-1$-polytope, just as in the 5-point case.

We can generalize to $A_{n;j}$ also, obtaining
\bea
&&A_{n;j}^{MHV}(n-j+2,...,n,1,...,n-j+1)\cr
&=&\sum_{\{\sigma\}_i\in OP(\{\a\},\{\b^T\})|_{j_{max}\in \{1,...,n-j+1\},l_{max}\in\{n-j+2,...,n-1\}}}
A_n^{MHV}(1,\{\sigma\}_i,n)\cr
&&\times \sum_{\{n-1,...,n-j+2\}\in \{\a\},\{\b\};j_{max}\in\{\a\},\{\b\}}(-)^{n_\b+j-1}M_n^{MHV}(\{\b\},1,\{\a\},n)\cr
&&\label{anjfinal}
\eea
where again we have pairs of $M_n^{MHV}$'s of different signs  and with $n-1$ common vertices adding up to give other 
closed polytopes (of $n+1$ vertices).

The new features with respect to the $A_{n;3}$ are as follows. The KK basis elements that we get are of a special type: 
If we take out $n-1,...,n-j+2$ from the amplitude, 
then the situation should be like the one for $n=3$, namely in the remaining permutation we go from 1 to a $j_{max}$ towards the right, and then towards
the left. But moreover, in $n-1,..., n-j+2$ we also have some ordering: some of them are in $\{\a\}$, some in $\{\b^T\}$, which means that 
$n-1,...,l_{max}+1$ is cyclic (i.e., towards the right), and $n-j+2,...,l_{max}$ is also cyclic (i.e., we change the direction of the cyclicity
at $l_{max}$). 

The number of terms multiplying a KK basis member is even, corresponding to having $j_{max}$ in $\{\a\}$ or $\{\b\}$ 
and any number of the $j-2$ terms $\{n-1,...,n-j+2\}$ in $\{\a\}$ and the rest in $\{\b\}$. They come in pairs, the pairs corresponding to 
$j_{max}$ being just before $n$ or just after, or otherwise one of the $\{n-1,...,n-j+2\}$ being either just before, or just after $n$, and 
the pairs as before having different sign.  
The sign of the terms is then simply $(-1)^{j-1+n_\b}$. In terms of polytopes, the two terms of different sign
correspond as before to polytopes with only a vertex differing between them, which means they again add up to another polytope with one more vertex.

As a simple application of this analysis, we note that (\ref{5pointKK}), (\ref{an3}) and (\ref{anjfinal}) show that the amplitudes 
$M_n$ come in alternating pairs. Each of 
these $M_n$ has leading IR singularity $1/\epsilon^2+{\cal O}(1/\epsilon)$, and therefore at one-loop $A_{n,j}$ has only a $1/\epsilon$ IR 
singularity. 

{\bf Polytope picture for the 6-point leading $NMHV$ amplitude}

The leading (planar) gluon amplitudes $A_{6;1}^{NMHV}$ for the split-helicity configuration are \cite{Bern:1994cg} 
\be
A_{6;1}^{NMHV}(1^+2^+3^+4^-5^-6^-)=\frac{c_\Gamma}{2}(B_1 W_6^{(1)}+B_2 W_6^{(2)}+B_3W_6^{(3)})\label{1lnmhv}
\ee
where $W_6^{(i)}$ are cyclic permutations of $W_6^{(1)}$, and $W_6^{(i+3)}\equiv W_6^{(i)}$, given in terms of box functions by 
\be
W_6^{(i)}=F_{6:i}^{1m}+F_{6:i+3}^{1m}+F_{6:2;i+1}^{2mh}+F_{6:2;i+4}^{2mh}\label{ws}
\ee
and the $F$'s are dimensionless boxes. We can write polytope interpretations for the $W_6^{(i)}$'s based on the fact that the $F$'s have 
polytope interpretation. Denoting for instance by $(4561(23))$ what was previously called $I(x_4,x_5,x_6,x_1(x_2,x_3))$, we write
\bea
W_6^{(1)}&=&(4561(23))+(1234(56))+(12(3)4(5)6)+(45(6)1(2)3)\cr
&\equiv&A_1+A_3+A_2+A_4\cr
W_6^{(2)}&=&(5612(34))+(2345(61))+(23(4)5(6)1)+(56(1)2(3)4)\cr
&\equiv&A_5+A_7+A_6+A_8\cr
W_6^{(3)}&=&(6123(45))+(3456(12))+(34(5)6(1)2)+(61(2)3(4)5)\cr
&\equiv&A_9+A_{11}+A_{10}+A_{12}
\eea
where the $A$'s are tetrahedra defined in the order they appear in the $W_6^{(i)}$ above, while for example for the 6-point MHV amplitude we have
\bea
A_{6;1}^{MHV}(123456)&=&A(123456)[(12(3)45(6))+(23(4)56(1))+(34(5)61(2))\cr
&&+(1234(56))+(2345(61))+(3456(12))\cr
&&+(4561(23))+(5612(34))+(6123(45))]\cr
&=&A(123456)[A_{13}+A_{14}+A_{15} +A_3+A_7+A_{11}+A_1+A_5+A_9] \cr
&&
\eea
where again the various $A$'s are defined in the order they appear. However, because of the spin coefficients of $W_6^{(i)}$, we cannot find a 
simple polytope interpretation for the subleading-color amplitudes. 

\subsection{Momentum twistor representation}

{\bf Momentum twistor representation for leading $N^kMHV$ amplitudes}

Instead, we can use a momentum twistor \cite{Hodges:2009hk,Hodges:2010kq}
representation for the $N^kMHV$ superamplitudes in order to find a simple formula for the subleading $N^kMHV$ 
amplitudes. 

The MHV tree-level color-ordered superamplitudes are given by the Nair formula \cite{Nair:1988bq}, 
a supersymmetric generalization of the Parke-Taylor formula \cite{Parke:1986gb,Berends:1987me}, 
\be
{\cal A}_{n,2}(12...n)=\frac{\delta^4(\sum_{i=1}^n\lambda_i\tilde\lambda_i)\delta^8(\sum_{i=1}^n\lambda_i\tilde \eta^i)}{\langle 12\rangle\langle
23\rangle...\langle n-1,n\rangle \langle n,1\rangle}\label{Nair}
\ee
where as usual $\langle ij\rangle\equiv \epsilon^{\a\b}\lambda_\a^{(i)}\lambda_\b^{(j)}$, $\tilde\eta$ is a spinor with an index $I=1,...,4$ for 
supersymmetries suppressed, and the $2$ in $A_{n,2}$ refers to R-charge, since 
the $N^kMHV$ amplitude has $m=k+2$ R-charge. 

The leading singularities of an amplitude are the discontinuities of the amplitude over the singularities where we put a maximum number of propagators 
on-shell, as explained in \cite{ArkaniHamed:2009dn}, where a conjecture for these leading singularities was proposed.

In terms of super-momentum twistors $Z_i$, the leading singularity of the (color-ordered, planar, i.e. leading) $N^kMHV$ super-amplitude is 
\cite{ArkaniHamed:2009vw,ArkaniHamed:2010gh}
\be
{\cal L}_{n,m}=\frac{\delta^4(\sum\lambda\tilde\lambda)\delta^8(\sum\lambda\tilde\eta)}{\langle12\rangle\langle 23\rangle...\langle
n1\rangle}\int\frac{d^{nk}{\cal D}}{Vol(Gl(2))}\frac{\prod_{\mu=1}^{k}\delta^{4|4}(\sum_{i=1}^n{\cal D}_{\mu 
i}Z_i)}{(12...k)(23...k+1)...(n12k-1)}={\cal L}_{n,2}\times {\cal R}_{n,k}\label{yangian}
\ee
where $k=m-2$. The prefactor ${\cal L }_{n,2}$ is the tree MHV amplitude (\ref{Nair}), 
and the integral ${\cal R}_{n,k}={\cal R}_{n,m-2}$ is Yangian invariant. 
This object is dual conformal {\em covariant}, only ${\cal R}_{n,k}$ being dual conformal invariant, and the tree amplitude is covariant.

The one-loop amplitudes of ${\cal N}=4$ SYM can be reduced to just boxes via the van Neerven and Vermaseren procedure, with some coefficients. 
The leading singularities also coincide with the coefficients of these box functions \cite{ArkaniHamed:2009dn}. For one-loop MHV, 
the coefficients of the boxes are known to be just the MHV tree amplitudes, agreeing with the result above. 

{\bf Subleading $N^kMHV$ amplitudes in momentum twistor space}

The planar (leading) color-ordered $N^kMHV$ amplitude is a sum of permutations of boxes with coefficients equal 
to the leading singularities,
\be
A_{n;1}(1...n)=\sum_\sigma {\cal L}_{n,k}(\sigma) I_{n;4}(\sigma)=\sum A^{MHV}_n(\sigma)R_{n;k}(\sigma)I_{n;4}(\sigma)
\ee
where $I_{n;4}$ are boxes. At 6-points, 
the permutations $\sigma$ combine such that we can organize the sum as a sum over cyclic permutations, with several boxes having the same coefficient
\cite{ArkaniHamed:2009dn}. 
For this coefficient we can factorize the tree MHV amplitude, which is cyclically invariant, so that it appears as a common factor 
\be
A_{6;1}(1...6)=A_6^{MHV}(1...6)\sum_{\lambda= \cyclic} R_{6;k}(\lambda)\sum_{\sigma/\lambda}I_{6;4}(\sigma)\label{a6twistor}
\ee
At higher $n$-point, the situation is slightly more complicated. The box diagrams are ordered in groups that can be cyclically permuted, 
for each group having a given formula for the residue, but unlike 6-point, the residue is not universal for all the 
groups \cite{ArkaniHamed:2009dn}. However, all 
the diagrams have still the external legs in the original order, which means, since the MHV tree amplitude
is cyclically invariant, that we can again factorize the MHV tree amplitude, obtaining for planar $N^kMHV$ amplitudes
\bea
A_{n;1}(1...n)&=&A_n^{MHV}(1...n)\sum_{{\rm groups\; of\; diagrams}}\sum_{\lambda= \cyclic} R_{n;k}(\lambda)\sum_{\sigma/\lambda}I_{n;4}(\sigma)\cr
&\equiv&A_n^{MHV}(1...n)M_{n;k}(1...n)\label{mnk}
\eea
which implicitly defines $M_{n,k}$. 

We now finally note that we have the same formula for $A_{n;1}(1...n)$ in terms of $A_n^{MHV}$ and $M_{n;k}$ from the previous section 
on polytopes, so we can apply the same calculations we used to obtain the $MHV$ $A_{n;j}$ in terms of $A_{n;1}$ in section 2. We just have to
change the definition of $M_{n;k}$ as in (\ref{mnk}), thus also drop the polytope interpretation of $M_{n,k}$. 
But otherwise the same (\ref{anjfinal}) found in the MHV case holds in the general $N^kMHV$ case as well, as can be seen from  
(\ref{an3}).

{\bf Application to the 6-point $NMHV$ amplitude}

For the superamplitude, we use an explicit form of the twistor formula (\ref{a6twistor}), 
doing the twistor space integrals over 
the 1-loop NMHV contours. The result is \cite{Kosower:2010yk,Drummond:2008bq}
\bea
A_{6;1}^{(1)NMHV}(123456)&=&\frac{a}{2}A_6^{(0)MHV}(123456)[(R_{413}+R_{146})W_6^{(1)}\cr
&&+(R_{524}+R_{251})W_6^{(2)}+(R_{635}+R_{362})W_6^{(3)}]\cr
&\equiv &A_6^{(0)MHV}(123456) M_6^{(1)NMHV}(123456)\label{6nmhv}
\eea
From the $R_{n;k}$ terms in (\ref{a6twistor}), one gets the sum of basic dual conformal invariant R-invariants $R_{j,j+3,j+5}$ above. 
Here the $R_{j,j+3,j+5}$ are given by
\bea
R_{rst}&=&-\frac{\langle s-1\; s\rangle\langle t-1\; t\rangle \delta^{(4)}(\Xi_{rst})}{x^2_{st}\langle r|x_{rt}x_{ts}|s-1\rangle\langle r|x_{rt}x_{ts}
|s\rangle\langle r|x_{rs}x_{st}|t-1\rangle \langle r|x_{rs}x_{st}|t\rangle }\cr
\Xi_{rst}&=&\sum_t^{r-1}\eta_i\langle i|x_{ts}x_{sr}|r\rangle +\sum_r^{s-1}\eta_i\langle i|x_{st}x_{tr}|r\rangle\cr
x_{st}&=&x_s-x_t=\sum_{i=s}^{t-1}p_i
\eea

As explained before, we can then perform the same combinatorics that led us to (\ref{an3}), 
just that now we use $M_6^{(1)NMHV}(123456)$ instead of the $M_6(123456)$. 

\section{Summary}

We have reviewed a number of features of subleading-color amplitudes
of ${\cal N}=4 $ SYM theory, a subject considerably less developed than
that of the leading (planar) amplitudes. Nevertheless this topic should
not be ignored if the structure of perturbative ${\cal N}=8$ supergravity
and its relationship to ${\cal N}=4 $ SYM theory is to be understood, as
non-planar graphs appear on an equal footing in ${\cal N}=8$ supergravity.

After presenting a detailed description of the IR divergences of ${\cal
N}=4$ SYM theory, we obtained the leading (and some subleading) IR
divergences of subleading-color amplitudes at $L$ loops, and tested these
against known exact results for one- and two-loop four-point functions.
These ideas applied to the one-loop five-point function led to a new KLT
relation, as well as possible new relations between ${\cal N}=4$ SYM
and ${\cal N}=8$ supergravity amplitudes.  A geometric interpretation
of the one-loop subleading and $N^kMHV$ amplitudes of ${\cal N}=4$
SYM was presented in the last section.

Since reformulations and extensions of known results frequently lead
to new insights, we advocate that continued study of subleading-color
amplitudes is likely to be fruitful. In particular, it would be important
for our understanding of the relation of ${\cal N}=4$ SYM to ${\cal
N}=8$ supergravity to extend (\ref{2k1}), (\ref{2k2}) to subleading
IR divergences, and to higher $n$-point functions. An example of the
latter is the speculative (\ref{type}) for $L=1$ and $n=5$. However,
(\ref{oneloopsugra}) and (\ref{twoloopre}) remind us that (\ref{type})
may not be the only way to relate the two theories, so that the subjects
discussed in this review should provide many opportunities for future
work.

{\bf Acknowledgments}

We have benefited from numerous insights and suggestions from Lance Dixon
throughout our work on the various topics of this review, for which we are
extremely grateful.
The research of 
S. G. Naculich is supported in part by the NSF under grant PHY-0756518.
The research of H. J. Schnitzer is  supported in part
by the DOE under grant DE--FG02--92ER40706. The research of H. Nastase is supported in part by CNPQ grant 301219/2010-9.

\providecommand{\href}[2]{#2}\begingroup\raggedright\endgroup

\end{document}